\documentclass[pre,amssymb,superscriptaddress,floatfix,twocolumn]{revtex4}
\usepackage{tikz}
\usetikzlibrary{snakes}
\usepackage{times}

\usepackage[colorlinks=true,linkcolor=blue,urlcolor=blue,citecolor=blue]{hyperref}

\usepackage{graphicx}
\usepackage{bm}
\usepackage{amsmath,color,amssymb,amsfonts}
\newcommand{\be}{\begin{equation}}
\newcommand{\ee}{\end{equation}}
\newcommand{\beq}{\begin{equation}}
\newcommand{\eeq}{\end{equation}}
\newcommand{\bea}{\begin{eqnarray}}
\newcommand{\eea}{\end{eqnarray}}
\newcommand{\rme}{\mathrm{e}}
\newcommand{\rmd}{\mathrm{d}}
\newcommand{\fig}[2]{\includegraphics[width=#1]{./#2}}

\newcommand{\nn}{\nonumber}
\newcommand{\half}{\frac12}
\renewcommand{\log}{\ln}
\newcommand{\ca}{\mathcal}
\newcommand{\eq}[1]{(\ref{#1})}
\newcommand{\Eq}[1]{Eq.~(\ref{#1})}
\newcommand{\Eqs}[1]{Eqs.~(\ref{#1})}
\newcommand{\w}{\nn\\&&}
\newcommand{\wa}{\nn\\&}
\newcommand{\lee}{\lefteqn}

\def\cF{{\mathcal{F}}}
\def\eps{\varepsilon}

\def\cD{\mathcal{D}}

\def\cO{\mathcal{O}}
\def\cS{\mathcal{S}}
\def\cP{\mathcal{P}}

\newcommand{\bal}{\begin{align}}
\newcommand{\eal}{\end{align}}

\def\fpt{t_{\rm FP}}

\def\posprop{P_+^{\mu,\nu}}
\def\brownposprop{P_+^{\mu}}

\renewcommand{\epsilon}{\varepsilon}

\newcommand{\checked}{}

\begin{document}

\bibliographystyle{KAY-hyper}

\title{Extreme Events for Fractional Brownian Motion with Drift: Theory and Numerical Validation}

\author{Maxence Arutkin}
\affiliation{UMR CNRS 7083 Gulliver, ESPCI Paris, 10 rue Vauquelin, 75005 Paris, France}
\author{Benjamin Walter}\affiliation{Department of Mathematics, Imperial College London, London SW7 2AZ, United Kingdom}
\author{Kay J\"org Wiese}
\affiliation{\mbox{Laboratoire de Physique de l'Ecole Normale Sup\'erieure, ENS, Universit\'e PSL, CNRS, Sorbonne Universit\'e,} 
\mbox{Universit\'e Paris-Diderot, Sorbonne Paris Cit\'e, 24 rue Lhomond, 75005 Paris, France.}}

\begin{abstract}
We study the first-passage time, the distribution of the maximum, and the absorption probability of fractional Brownian motion of Hurst parameter $H$  with both a linear and a non-linear drift. The latter appears naturally when applying non-linear variable transformations. Via a perturbative expansion in $\epsilon = H-1/2$, we  give the first-order corrections to the classical result for Brownian motion analytically. Using a recently introduced  adaptive bisection algorithm, which is much more efficient than the standard Davies-Harte algorithm,  we test our predictions  for the first-passage time on grids of  effective  sizes   up to $N_{\rm eff}=2^{28}\approx 2.7\times 10^{8}$ points. The  agreement between theory and simulations is  excellent, and by far exceeds in precision what can be obtained by scaling alone. 
\end{abstract}

\maketitle

\section{Introduction}
Understanding the extreme-value statistics of random processes is important in a variety of contexts. Examples are records  \cite{MajumdarSchehrWergen2012}, e.g.\ in climate change \cite{WergenBognerKrug2011}, equivalent to depinning \cite{LeDoussalWiese2008a},   in quantitative trading \cite{RejSeagerBouchaud2017}, or for earthquakes \cite{ShomeCornellBazzurroCarballo1998}. While much is known for Markov processes, and especially for Brownian motion \cite{RednerBook,GumbelBook,FellerBook,FellerBook2,BorodinSalminen2002,BertoinBook,Wiese2019}, much less is known for correlated, i.e.\ non-Markovian processes, of which   fractional Brownian motion (fBm) is the simplest   scale-free version  \cite{NourdinBook,Sottinen2001,Sinai1997,MandelbrotVanNess1968,Krug1998,DiekerPhD,DiekerMandjes2003,Aurzada2011}.

FBm is important as it successfully models  a variety of natural processes \cite{DecreusefondUstAOEnel1998}: a tagged particle in   single-file diffusion ($ H\,{=}\,0.25 $) \cite{KrapivskyMallickSadhu2015,SadhuDerrida2015}, the integrated current in diffusive transport ($ H\,{=}\,0.25 $) \cite{SadhuDerrida2016}, polymer translocation through a narrow pore ($ H\,{\simeq}\,0.4 $) \cite{ZoiaRossoMajumdar2009,DubbeldamRostiashvili2011,PalyulinAlaNissilaMetzler2014}, anomalous diffusion \cite{BouchaudGeorges1990}, values of the log return of a stock  ($H\,{\simeq}\,0.6\; {\rm to }\; 0.8 $) \cite{Peters1996,CutlandKoppWillinger1995,BiaginiHuOksendalZhang2008,Sottinen2001}, hydrology ($H\,{\simeq}\,0.72\;{\rm to}\;0.87 $) \cite{MandelbrotWallis1968}, a tagged monomer in a polymer   ($ H\,{=}\,0.25$) \cite{GuptaRossoTexier2013}, solar flare activity ($H\,{\simeq}\,0.57\;{\rm to}\;0.86$) \cite{Monte-MorenoHernandez-Pajares2014}, the price of electricity in a liberated market ($ H\,{\simeq}\,0.41 $) \cite{Simonsen2003}, telecommunication networks ($H\,{\simeq}\,0.78\;{\rm to}\;0.86$) \cite{Norros2006}, telo\-meres inside the nucleus of human cells ($H\,{\simeq}\,0.18\;{\rm to}\;0.35$) \cite{BurneckiKeptenJanczuraBronshteinGariniWeron2012}, or diffusion inside crowded fluids ($ H\,{\simeq}\,0.4 $) \cite{ErnstHellmannKohlerWeiss2012}. 

Recently, first-passage times of fBm have been investigated    \cite{JeonChechkinMetzler2011,JeonChechkinMetzler2013,GuerinLevernierBenichouVoituriez2016,DelormeWiese2016,DelormeThesis,DelormeWiese2015}.  Due to the non-Markovian nature of the process,  translating these results to a fBM with drift is far from trivial, and even properly estimating the drift for $H<1/2$ is a challenge \cite{Es-SebaiyOuassouOuknine2009}. To our knowledge, no anaytical result for a fBm with drift are known.   It is this gap we intend to fill here.  


As is discussed later, 
apart from a {\em linear drift},   a {\em  non-linear drift}  may appear as well, leading us to consider the  process, 
\bea 
\label{1}
z_t &:=& x_t + \mu t + \nu t^{2H} \ .
\eea
Here $x_t$ is a standard fractional Brownian motion (fBm) with 
mean and variance
\bea
\left< x_t \right> &=& x_{0 } = 0 \ ,\\
\left< x_{t_1} x_{t_2}\right> &=& |t_1|^{2H}+|t_2|^{2H}-|t_1-t_2|^{2H}\ .
\eea
The parameter $H$ is the Hurst parameter. 
Since fBm is a Gaussian process, the above equations uniquely and completely specify it. 
Taking a derivative w.r.t.\ both $t_1$ and $t_2$ shows that the increments of the process are correlated, 
\be
\left< \dot x_{t_1} \dot x_{t_2}\right> = 2H(2H-1) |t_1-t_2|^{2H-2}\ .
\ee
Correlations are positive for $H>1/2$, and negative for $H<1/2$. The case $H=1/2$ corresponds to Brownian motion, with uncorrelated increments.

The parameters $\mu$ and $\nu$ are the strength of linear and non-linear drift. While linear drift is a canonical choice, non-linear drift   appears as a consequence  of     non-linear variable transformations. As an example, consider the process
\be
y_{t} := \rme^{z_{t}}\ .
\ee
The exponential  transformation appears quite often, be it in the Black-Sholes theory of the stock market where the logarithm of the portfolio price is treated as a random walk  \cite{BlackScholes1973,CutlandKoppWillinger1995,BouchaudPotters2009}, be it in non-linear surface growth of the Kardar-Parisi-Zhang universality class \cite{KPZ,Wiese1998a,JanssenTauberFrey1999}, where the transformation is known as the Cole-Hopf transformation \cite{Hopf1950,Cole1951}, or in the evaluation of the Pickands constant \cite{DebickiKisowski2008,DelormeRossoWiese2017,HaanPickands1986,Harper2014,Michna2009,Pickands1969,Pickands1971,Pickands1975}.
Like any non-linear transform, this generates an effective drift known from It\^{o}-calculus. Computing the average of $y_t$ gives
\bea
\left< y_{t} \right> &=& \left<  \rme^{z_{t}} \right> =  \exp\!\left({\left< z_{t} \right> + \frac12 \left[\left< z_{t} ^{2} \right> - \left< z_{t} \right>^2\right]}\right) \nn\\
&=&  \exp\!\Big( \mu t + \left[ \nu +1\right]  t^{{2H}} \Big) \ .
\eea 
Thus even if   initially there is no nonlinear drift, it is   generated by non-linear transformations. For this reason, we include it into our model.

While for Brownian motion, equivalent to $H=\frac12$, many results can be obtained analytically \cite{RednerBook,GumbelBook,FellerBook,FellerBook2,BorodinSalminen2002,BertoinBook,Wiese2019}, for fBm much less is known. Recently, some of us  developed a framework \cite{WieseMajumdarRosso2010} for a systematic expansion in 
\be
\epsilon  := H -\frac 12\ .
\ee 
It has since successfully been applied to obtain the distribution of the maximum and minimum of an fBm \cite{DelormeWiese2015,DelormeWiese2016}, to fBm bridges \cite{DelormeWiese2016b},
  evaluation of the Pickands constant \cite{DelormeRossoWiese2017},   the 2-sided exit problem \cite{Wiese2018}
and 
  the generalization of the three classical arcsine laws
\cite{SadhuDelormeWiese2017}. It is also known that the fractal dimension of the record set of an fBm is $d_{\rm f}=H$ \cite{BenigniCoscoShapiraWiese2017}.

This article is organized into four sections, the introduction,  theory in section \ref{s:Theory}, and numerics in section \ref{s:Numerics}, followed by conclusions in section \ref{s:Conclusion}.

\begin{table}
\begin{tabular}{|c|c|}
\hline
 \bf P & \mbox{probability} \\
 \hline
$ ~~P = \partial_{x } \bf P ~~$ & ~~\mbox{probability density in $x$}~~\\ 
\hline
$~~ \mathbb P = \partial_{t} \bf P ~~$ & ~~\mbox{probability density in $t$}~~\\ 
\hline
$~~ \ca P = \partial_{y} \bf P~~$ &~~ \mbox{probability density in $y$}~~\\
\hline
\end{tabular}
\caption{Notations used for probabilities and their various densities.}
\label{t:notations}
\end{table}
\section{Theory}
\label{s:Theory}{
In this section, we find the probability distribution of first-passage times and running maxima of fBm  with linear and non-linear drift by way of a perturbation expansion around   simple Brownian motion. The key result of this section is the scaling function \eqref{P(y|m)-final} which together with the auxiliary functions defined in Eqs.~\eqref{F1-def}, \eqref{Fmu} and \eqref{Fnu} gives the distribution of first-passage times. The majority of this section is devoted to deriving these results.
\subsection{Scaling  dimensions}
\label{sec:scaling}
Before developing the perturbation theory, we consider the scaling dimensions involved. This will be useful for later discussion of the scaling functions.
For fBm as defined in \Eq{1}, there are four dimension-full quantities, $x$, $t$, $\mu$, and $\nu$. Scaling functions will thus depend on three scaling variables, which we now identify. We start with the terms without drift: 
\be
x\sim t^H ~~\Longleftrightarrow~~ t \sim x^{\frac1 H}\ ,
\ee
where the tilde means ``same scaling dimension''. 
Thus (without drift), any observable $\mathfrak O(x,t)$ can be written as 
\be\label{y}
\mathfrak O( x,t) = x^{{\rm dim}_x({\mathfrak O})} f_{\mathfrak O} ( y ) \ , \qquad  y :=\frac{x}{\sqrt2 t^H}
\ .
\ee
The variable $y$ is dimension free.  
In presence of a linear drift, one has
\be
x \sim \mu t ~~\Longleftrightarrow~~\mu \sim \frac x t \sim x^{1-\frac 1H} \sim t^{H-1}
\ .
\ee
Thus the   combination $u = \mu x^{\frac1H-1}$ is dimension free, as  is 
$\tilde u := u^{\frac{H}{1-H}} = \mu ^{{\frac{H}{1-H}}} x$.
For non-linear drift, we have
\be
x\sim \nu t^{2H}~~\Longleftrightarrow~~\nu \sim \frac x {t^{2H}} \sim \frac1 x \sim \frac1 {t^{H}}\ .
\ee
Another scaling variable therefore is 
$
v = \nu x. $ In conclusion, any observable $\ca O $ can, in generalization of \Eq{y}, be written as
\bea\label{12}
\mathfrak O( x,t,\mu,\nu) &=& x^{{\rm dim}_x({\mathfrak O})} f_{\mathfrak O} ( y,u,v ) \ , \\
 y &=& \frac{x}{\sqrt2 t^H}\ ,\\
 u =  \mu x^{\frac1H-1},  &&  \mbox{or\quad} \tilde u  = \mu ^{{\frac{H}{1-H}}} x \ ,\\
 v &=& \nu x\ .
\label{15bis}
\eea

\subsection{The first-passage time}
The central result of our work is a perturbative expression of the first-passage-time density of fBM with linear and nonlinear drift as introduced in Eq.~\eqref{1}. The first-passage time $\fpt$ is   defined as
\be
\fpt(m) := \inf_{t>0} \left\{ t, z_{t} \le 0 | z_{t=0} = m \right\}\ ,
\label{def:first_passage_time}
\ee
where $m$ is the starting point of the process $z_t$, and  $m > 0$. 
The first-passage-time density for Brownian motion with (linear) drift, see e.g.\ \cite{RednerBook}, and rederived below in \Eq{p-abs}, is 
\bea
\label{17} 
\mathbb P_{0}(\fpt(m) = t)=  \frac{m }{2 \sqrt{\pi } t^{3/2}} \rme^{-\frac{1}{2}\!  \big( \frac{m}{\sqrt{2t}}+\frac\mu2 \sqrt{2t} \big)^2}\ .
\eea
This density in time is most naturally expressed in terms of the scaling variable $y$ introduced in  \Eq{y}, and which for Brownian motion ($H=1/2$) reads}
\be 
y = \frac{m}{\sqrt{2 t}}\bigg|_{t=t_{\rm FP}(m)}  \ .
\ee
For  Brownian Motion  the probability distribution of $y$ takes the simple form
\begin{align}
\label{19}
\cP_{0}(y; \mu) &= \sqrt{\frac{2}{\pi}} \, \rme^{ - \cF_0(y;\mu)}\ ,\\
\cF_0(y;\mu) &=  \half  \left(y + \frac{\mu}{2}\frac{m}{y}\right)^2\ .
\end{align}
Note that the measure is $\rmd t$ in \Eq{17} (density in time), whereas in \Eq{19} it is $\rmd y$ (density in $y$).  To avoid confusion, we   use distinct symbols for probabilities $\mathbf P$, densities $\mathbb P$ in time $t$, densities $\cal P$  in $y$, and densities $P$ in space $x$, independent of the actual choice of variables. This is summarized in table \ref{t:notations}.

We   introduced the scaling function $\cF_0$.
Below  we compute its  corrections to first order in $\varepsilon$, leading to a correction of the first-passage density in $y$, 
\begin{align}
 {\cP(y; \mu, \nu) } = \frac{ y^{\frac1H - 2}}{\sqrt{2 \pi}} \rme^{-  \cF_0(y;\mu, \nu) -\eps \delta \cF(y;\mu,\nu)} + \cO(\eps^2)\ .
\end{align}
The result is given in Eqs.~\eq{74}-\eq{P(y|m)-final}. 
Two comments are in order: {\it(i)}
the exponential resummation   is chosen for better convergence for larger $\epsilon$, as discussed in \cite{Wiese2018}, section IV.C;
{\it (ii)}
the distribution of first-passage times is related to the distribution of maxima. 

Readers wishing to skip ahead will find the 
    function $\delta \cF$      evaluated using path-integral methods, described in section \ref{s:The path-integral of a fBm with drift}. For the explicit result,  see section \ref{s:Scaling and corrections from the diffusion constant, final result}. A confirmation by numerical simulations is shown in section \ref{s:Simulation results}. 

\subsection{Summary of calculations to be done}
In order to calculate the first-passage-time distribution, we consider the process $z_t > 0$ in the presence of an absorbing boundary condition at $z = 0$ and restrict ourselves to  $z_t > 0$.
The transition probability density of the process $z_t$ to pass from $z_0>0$ to $z_1>0$ in   time $t$, without being absorbed at $z=0$ is denoted   $\posprop(z_0, z_1;t)$.  The probability density of first-passage times $\mathbb P \big(\fpt(m)=t\big)$  can then be obtained as
\begin{align} 
\mathbb P(\fpt(m) = t) = \left. \partial_{z_1}  \posprop(m, z_1,t)\right|_{z_1 = 0}.
\label{eq:fpt_is_derivative}
\end{align}
This relation holds since the derivative on the right-hand-side picks out those trajectories which assume $z_t = 0$ at time $t$ for the first time. The general strategy of this work is   to   compute $\left.\partial_{z_1}  \posprop(m, z_1,t)\right|_{z_1 = 0}$ and its perturbative corrections using path-integral methods. In the subsequent section \ref{s:Simple Brownian Motion: First-passage time and absorption probability}, we discuss the reference point of our expansion, simple Brownian motion. In  section \ref{s:The path-integral of a fBm with drift}, we introduce a perturbative expansion around Brownian motion, based on a path-integral formalism.  This yields a diagrammatic expansion (section \ref{s:Diagrammatic expansion}), with three diagrams, listed in section \ref{s:Diagrams to be evaluated},   evaluated in sections \ref{s:G1} 
 to \ref{s:Gbeta}, and regrouped in section \ref{s:Combinations}. The final result is given in section \ref{s:Scaling and corrections from the diffusion constant, final result}. Contrary to the drift-free case, not all processes are absorbed, as is discussed in section \ref{s:Absorption probability}. Relations between the different probability densities are discussed in section \ref{s:remark}, followed by an analysis of the tail of these distributions in section \ref{s:tails}.
Numerical checks are presented in section \ref{s:Numerics}, followed by conclusions in section \ref{s:Conclusion}.

\subsection{Simple Brownian Motion: First-passage time and absorption probability}
\label{s:Simple Brownian Motion: First-passage time and absorption probability}
The perturbation theory is an expansion around simple Brownian motion. This base point is considered here. By setting $H = \frac12$  and $\nu = 0$ in \Eq{1}, we obtain simple Brownian motion with drift. For this process, we compute \emph{(i)} the positive transition probability and \emph{(ii)} the absorption probability.

The   transition probability of simple Brownian motion $\brownposprop$ (to alleviate our notations, we do not put an index $0$ to indicate Brownian motion, since $P_{+}$ is not used for fBm), the probability to pass from $z_0$ to $z_1$ within time $t$ without crossing the line  $z \equiv 0$, satisfies the associated Fokker-Planck equation
\begin{equation}
\partial _t  \brownposprop(z_0,z_1,t) = \partial_{z_1}^2\brownposprop(z_0, z_1, t) - \mu \partial_{z_1} \brownposprop(z_{0},z_1,t)
\ .
\end{equation}
with appropriate absorbing boundary condition at $z \equiv 0$. Its solution is given by  the mirror-charge solution
\bea\label{3}
\brownposprop(z_0,z_1,t)&=&\frac{1}{\sqrt{4\pi t}}\left(\rme^{-(z_1-z_0)^2/4t} -\rme^{-(z_1+z_0)^2/4t}\right)\nn\\
&&  \times\rme^{\frac\mu 2 (z_1-z_{0}) -\frac{\mu^{2}t}4}
\ ,
\eea
satisfying the initial condition
\begin{equation}\label{f1}
\brownposprop (z_{0},z_{1},t=0) = \delta (z_{0}-z_{1})\ .
\end{equation}
It is useful to   consider its Laplace-transformed version. 
We define the Laplace transform of a function $f (t)$, with
$t\ge 0$ as 
\begin{equation}\label{f56-bis} 
\tilde f (s) 
:=\mathcal{L}_{t\to s }
\left[ f (t) \right]  =  \int_{0}^{\infty}\rmd t\,  \rme^{-s t} f (t) \ .
\end{equation}
This yields 
\begin{equation}\label{16-bis}
\tilde P_+^{\mu }(z_0,z_{1},s) = \rme^{\frac\mu 2 (z_{1}-z_0)} \tilde P_+\Big(z_0,z_{1},s+\frac{\mu^{2}}4\Big) \ , \end{equation}
where the drift-free propagator reads
\be\label{16-cis}
\tilde  P_+(z_0,z_{1},s) = \frac{\rme^{-\sqrt{s }  \left(z_0-z_1\right)}-\rme^{-\sqrt {s } \left(z_0+z_1\right) }}{2 \sqrt{{s }} } \ .
\ee
The Laplace transform  $\tilde{\mathbb{P}}(m,s)$ of the first-passage-time probability density, following  Eq.~\eqref{eq:fpt_is_derivative}, equals the probability to go close to the boundary, and there being absorbed for the first time, 
\begin{align}\nn
\tilde{\mathbb{P}}(m,s) &:= \int_0^{\infty} {\rm d}t \, e^{-s t}  {\mathbb{P}}(\fpt(m) = t) \\
\nn
& = \left. \partial_{z_1} \tilde P_+^\mu (m,z_{1},s) \right|_{z_1 = 0} \\
&=   \rme^{-\frac\mu 2 m} \rme^{-m\sqrt {s+\mu^{2}/4}}\ .
\label{32}
\end{align}
Its inverse Laplace transform is the first-passage-time probability density
\bea
\label{p-abs}
 \mathbb{P}(\fpt(m) = t)&=&  \rme^{-\frac\mu 2 m -\frac{\mu^{2}}4 t}\frac{m \rme^{-\frac{m^2}{4 t}}}{2 \sqrt{\pi } t^{3/2}}\ ,
\eea
confirming the result in Eq.~\eqref{17}. The total (time integrated) absorption probability   is 
\bea\label{p-abs-2}
\mathbf P_{\rm abs}(m) &=&  \tilde{\mathbb{P}}(m,s=0) \nn\\
&=&  \rme^{-\frac\mu 2 m} \rme^{-\frac{|\mu|}2  m} = \left\{ \begin{array}{ccc}
\rme^{-\mu m} &,&~\mu>0
\\
1 &,& ~\mu\le 0
\end{array}
\right. 
\ .\qquad
\eea
In what follows, we present perturbative corrections of these results for $\eps \neq 0$.
%

\subsection{The path-integral of a fBm with drift}
\label{s:The path-integral of a fBm with drift}
The technology developed in  \cite{WieseMajumdarRosso2010,DelormeWiese2016,Wiese2018}   uses a path-integral   to describe   fBM. Since $z_t$ is Gaussian, its path-probability measure on a finite interval $[0,T]$ is 
\begin{align}
{\bf P}[z_t] = \exp\left( -\cS[z_t;\mu,\nu]\right)\ ,
\end{align}
where $\cS[z_t;\mu,\nu]$ is an action quadratic in   $z_t$.
Without drift  ($\mu = \nu = 0$), the action for a  fBM   to   order   $\eps$ is \cite{WieseMajumdarRosso2010,DelormeWiese2016,Wiese2018} 
\begin{align}
\label{B1}
&\cS[z_t;\mu= \nu=0] \\
&= \int_{0}^{T} \rmd t\,\frac{ \dot z_{t}^2} {4 D_{\epsilon}} - \frac{\epsilon}2 \int_{\tau}^T \rmd t_2 \int_0^{t_2-\tau} \rmd t_1 \frac{\dot z_{t_1} \dot z_{t_2}  }{|t_1-t_2|} \nonumber  \ .
\end{align}
The action consists of a local part, corresponding to simple Brownian motion, and a non-local part, proportional to $\epsilon$.
The idea behind the perturbative expansion is that Brownian motion (as given by the first term) samples the whole phase space of fBm, albeit with the wrong probability measure. Our perturbation theory corrects this, by   weighing each path with the second term in \Eq{B1}. 
This   implies that the absorbing boundary conditions at the origin are properly taken into account, and that  observables as the absorption current, which are given by local operators, remain valid. 
%
%
 For regularity, a short-distance cutoff $|t_1-t_2|>\tau$ is introduced in the last integral, which is reflected in the diffusion constant \cite{DelormeWiese2016}
\be
D_{\epsilon} =  2H \tau^{2H-1} = (1+ 2\epsilon) \tau^{2\epsilon} = (\rme \tau)^{2\epsilon} + \ca O(\epsilon^2)
\ .
\ee
Let us now insert the   definition (\ref{1}) into the action \eq{B1}. The reason to   proceed this way  is that the method of images on which our further calculation relies   works in terms of  $x_t$ as defined in \Eq{1}, but not $z_t$.  After some algebra  we arrive  at the action for an arbitrary drift 
\bea
{\cal S}[z_t] 
&=&  \int_{0}^{T} \rmd t\,\frac{ \dot z_t^2} {4 D_{\epsilon}} 
\w
+\int_{0}^{T} \rmd t \frac{\epsilon}{2}  \dot z_t \left[(\mu {+}\nu )
   \log \left(\frac{t (T{-}t)}{\tau ^2}\right)-2
   \nu  \log \left(\frac t{ \tau}\right)\right]   
\w   
     - \frac{\epsilon}2 \int_{\tau}^{T}\rmd  t_2\int_{0}^{t_2-\tau}\rmd  t_1  \frac{\dot z_{t_1} \dot z_{t_2}  }{|t_1-t_2|}   \nn\\
&& - \frac{z_{T}-z_{0}}{2 }\Big[{ \frac{\mu}{D_{\epsilon}}+\nu}\Big]+\frac{T}{4}  (\mu +\nu )^2
\w
+\frac{T}{2} 
   \epsilon  \left(\nu ^2-\mu ^2\right) \log
   (T)+\ {\cal O}(\epsilon^2)\ .
\label{77}
\eea
Some checks are in order.  In absence of absorbing boundaries,  the exact free propagator reads
\bea
\!\! P^{\mu,\nu}(0,z ,T) &=&\frac1{2\sqrt \pi T^H}\rme^{-\frac{(z-\mu T-\nu T^{2 H})^2}{4T^{2H}}} \nn\\
& =&\frac1{2\sqrt \pi T^H}\exp\bigg({-}\frac{z^2}{4 T^{2H}}  + \frac{ z}2 \Big[\nu {+} \mu T^{-2\epsilon}\Big]
\w
~~~~~~~~~~~~~~~~~~~~~~~~~  {-} \frac{ T }{4}   \Big[\nu T^\epsilon {+}\mu T^{- \epsilon}\Big]^{2}\bigg) 
 .~~~~~~~\eea
Since the above formalism has variables $\dot z$ only, the term $\sim z^2$ is given by the drift-free perturbation theory. We can further check that if we replace in the action  $\dot z(t)$ by its ``classical trajectory'', i.e. $\dot z(t)\to [z(T)-z(0)]/T$, then both the normalization and the drift term agree with the exact propagator.

Let us specify \Eq{77} to the two cases of interest:
For a fBm with {\em linear drift} as given in Eq.~(\ref 1) with $\nu=0$, we have 
\bea
{\cal S}_{\nu = 0}[z_t ]
&=&  \int_{0}^{T} \rmd t\,\frac{ \dot z_t^2} {4 D_{\epsilon}} - \frac\mu{2 D_{\epsilon}} (z_T-z_0)  + \frac {T^{1-2 \epsilon}}4 \mu^2
\w
- \frac{\epsilon}2  \int_{\tau}^{T}\rmd  t_2\int_{0}^{t_2-\tau}\rmd  t_1  \frac{\dot z_{t_1} \dot z_{t_2}  }{|t_1-t_2|} 
\w
 +  \frac{\epsilon\mu}{2}   \int_{0}^{T} \rmd t\, \dot z_{t}  \ln \left(\frac{[T-t]t}{\tau^2}\right)   +\ca O(\epsilon^2)  \ .~~~~~~~
\eea
For a fBm with {\em non-linear drift} as given in Eq.~(\ref 1) with $\mu=0$, we have 
\bea
{\cal S}_{ \mu = 0}[z]
&=&  \int_{0}^{T} \rmd t\,\frac{ \dot z_t^2} {4 D_{\epsilon}} - \frac{\nu}{2} (z_T-z_0)  + \frac {T^{1+2 \epsilon}}4 \nu^2
\w
- \frac{\epsilon}2  \int_{\tau}^{T}\rmd  t_2\int_{0}^{t_2-\tau}\rmd  t_1 \frac{\dot z_{t_1} \dot z_{t_2}  }{|t_1-t_2|} 
\w
 +  \frac{\epsilon\nu}2   \int_{0}^{T} \rmd t\, \dot z_{t}  \ln \left(\frac{T-t}{t}\right)    +\ca O(\epsilon^2) \ .~~~~~~~
\eea
Note the appearance of the diffusion constant in the ``bias'' (Girsanov) term $z_T-z_0$ for a linear drift, and its absence for a non-linear drift.

To simplify the notation, we introduce 
\begin{align}
\cS_0[z_t] =   \int_{0}^{T} \rmd t\,\frac{ \dot z_t^2} {4}
\end{align}
as a shorthand   for the   Brownian action around which perturbation theory expands. 
The drift (Girsanov) term is $\rme^{-\ca S_{\rm d}}$, with 
\be
\ca S_{\rm d}[z] = \frac{ z_{0}-z_{T}}2\left( \frac\mu{D_\epsilon}{+}\nu\right) +\frac T4 \left( \mu  T^{-  \epsilon} {+}\nu  T^{\epsilon}\right)^{2}  \ .
\ee
Further,    define  (valid at leading order in $\epsilon$) 
\bea\label{14a}
\alpha&:=& \mu-\nu \ , \qquad \beta := \mu+\nu\ ,\\
\mu &=& \frac{\alpha+\beta}2\ , \qquad \nu =\frac{\beta -\alpha}2\ . \label{14b}
\eea
This  simplifies the drift terms   in the action to 
\begin{align}
\label{S-alpha}
{\cal S}_{\alpha}[z_t]
&:=  \frac{1 }{ 2}    \int_{0}^{T}
\rmd t\, \dot z_{t}  \ln \left(\frac {t}{\tau}\right) \ ,\\ 
\label{S-beta}
{\cal S}_{\beta} [ z_t]
&:= \frac{1 }{ 2}    \int_{0}^{T}
\rmd t\, \dot z_{t}  \ln \left(\frac {T-t}{\tau}\right) \ .
\end{align}
Finally, the drift-independent perturbative correction containing the non-local interaction reads
\be\label{S-1}
\cS_{1}[z_t]=    \frac{1}2 \int_{\tau}^{T}\rmd  t_2\int_{0}^{t_2-\tau}\rmd  t_1  \frac{\dot z_{t_1} \dot z_{t_2}  }{|t_1-t_2|}   
\ .
\ee
In   these notations,  the action to order $\epsilon$ reads
\be
\cS[z_t; \mu, \nu] =  \frac{\cS_0}{D_{\epsilon}}  + \ca S_{\rm d}- \varepsilon \left( \cS_1 -\alpha  \cS_{\alpha} -\beta \cS_{\beta} \right).
\ee
Perturbation theory takes   place in the three interaction-terms proportional to $\eps$, plus an additional contribution due to $D_\epsilon$. The bare result \Eq{16-bis} of transition probabilities of fBM  will thus be corrected by three different terms corresponding to the three interaction terms $\cS_{\alpha}, \cS_{\beta}$ and $\cS_1$, plus a correction from $D_\epsilon$. The (diagrammatic) rules for computing these corrections are outlined in the next section.

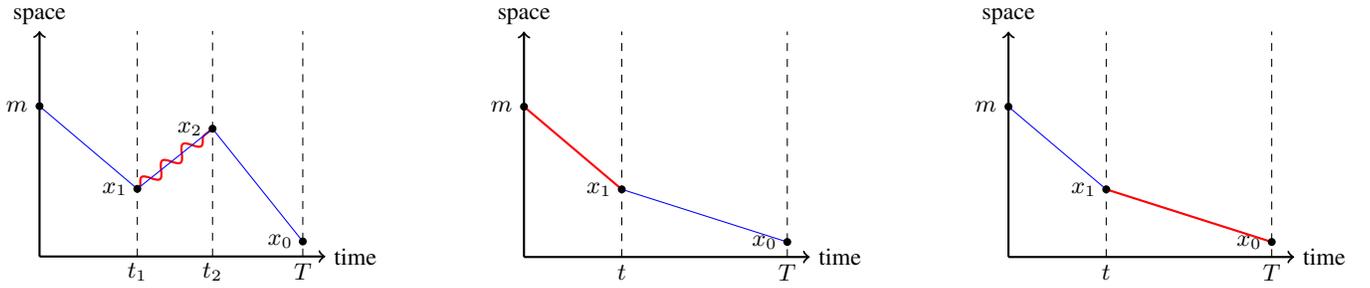
\begin{figure*}
{{\begin{tikzpicture}
\draw [->,thick] (0,0) -- (3.8,0);
\draw [->,thick] (0,0) -- (0,3);
\draw [dashed] (3.5,0) -- (3.5,3);
\draw [dashed] (2.3,0) -- (2.3,3);
\draw [dashed] (1.3,0) -- (1.3,3);
\node (m1) at  (0,2) {$\hspace{-6mm}m$};
\node (x1) at (1.3,0.9) {$\hspace{-6mm}x_1$};
\node (x2) at (2.3,1.7) {$\hspace{-6mm}x_2$};
\node (x0) at (3.5,0.2) {$\hspace{-6mm}x_0$};
\draw [snake=snake,red,thick] (x1) -- (x2);
\draw [blue] (m1) -- (x1);
\draw [blue] (x1) -- (x2);
\draw [blue] (x2) -- (x0);
\fill (m1) circle (1.5pt);
\fill (x0) circle (1.5pt);
\fill (x1) circle (1.5pt);
\fill (x2) circle (1.5pt);
\node (t1) at  (1.3,-0.2) {$t_1$};
\node (t2) at (2.3,-0.2) {$t_2$};
\node (T) at (3.5,-0.2) {$T$};
\node (time) at (4.2,0) {time};
\node (space) at (0,3.2) {space};
\end{tikzpicture}}}\hfill
{{\begin{tikzpicture}
\draw [->,thick] (0,0) -- (3.8,0);
\draw [->,thick] (0,0) -- (0,3);
\draw [dashed] (3.5,0) -- (3.5,3);
\draw [dashed] (1.3,0) -- (1.3,3);
\node (m1) at  (0,2) {$\hspace{-6mm}m$};
\node (x1) at (1.3,0.9) {$\hspace{-6mm}x_1$};
\node (x0) at (3.5,0.2) {$\hspace{-6mm}x_0$};
\draw [red,thick]  (m1) -- (x1);
\draw [blue] (x1) -- (x0);
\fill (m1) circle (1.5pt);
\fill (x0) circle (1.5pt);
\fill (x1) circle (1.5pt);
\node (t1) at  (1.3,-0.2) {$t$};
\node (T) at (3.5,-0.2) {$T$};
\node (time) at (4.2,0) {time};
\node (space) at (0,3.2) {space};
\end{tikzpicture}}}\hfill
{{\begin{tikzpicture}
\draw [->,thick] (0,0) -- (3.8,0);
\draw [->,thick] (0,0) -- (0,3);
\draw [dashed] (3.5,0) -- (3.5,3);
\draw [dashed] (1.3,0) -- (1.3,3);
\node (m1) at  (0,2) {$\hspace{-6mm}m$};
\node (x1) at (1.3,0.9) {$\hspace{-6mm}x_1$};
\node (x0) at (3.5,0.2) {$\hspace{-6mm}x_0$};
\draw [blue]  (m1) -- (x1);
\draw [red,thick] (x1) -- (x0);
\fill (m1) circle (1.5pt);
\fill (x0) circle (1.5pt);
\fill (x1) circle (1.5pt);
\node (t1) at  (1.3,-0.2) {$t$};
\node (T) at (3.5,-0.2) {$T$};
\node (time) at (4.2,0) {time};
\node (space) at (0,3.2) {space};
\end{tikzpicture}}}
\caption{Graphical representation of the
path-integral for  diagram ${\mathbb G}_1(m,t)$ (left, expectation of $\ca S_1$), ${\mathbb G}_{\alpha}(m,t)$ (middle,  expectation of $\ca S_\alpha$),  and  ${\mathbb G}_{\beta }(m,t)$ (right,  expectation of $\ca S_\beta$).
The wiggly line in the first diagram represents the interaction proportional to $1/(t_2-t_1)$. The red lines in the second and third diagram contain a log of the corresponding time difference, $\ln(t/T)$ for the first, and $\ln\big((T-t)/T\big)$ for the second. 
}\label{f:G1+G2+G3}
\end{figure*}

\medskip

\subsection{Diagrammatic expansion}
\label{s:Diagrammatic expansion}

The central aim of this work is to calculate the first-passage-time density. This is done by taking the derivative of the survival transition density at its endpoint (cf.~\Eq{eq:fpt_is_derivative}). The latter is obtained perturbatively by evaluating a path-integral over the action defined previously.

\bea
\mathbb P^{\mu,\nu}(m,t) &:=&   \partial_{z_1}   P_{+,\epsilon}^{\mu,\nu}(m,z_1,t)\Big|_{z_{1}=0}
\nn\\
&\equiv& \lim_{ z_{1}\to 0 }\frac1 {z_{1}}P_{+,\epsilon}^{\mu,\nu}(m,z_1,t)\ .
\label{def:G_as_deriv}
\eea
Here we introduced $P_{+,\epsilon}^{\mu,\nu}(m,z_1,t)$
\begin{align}
&P_{+,\epsilon}^{\mu,\nu}(m, z_1,t):= 
\int_{z_{0}= m}^{z_t = z_1} \cD[z_t] \Theta(z_t) \exp\left( - \ca S \right)\ ,
\end{align}
 the probability of a path $z_t$ to pass from $m$ to $z_1$ within time $t$ without being absorbed at $z=0$ (cf.~Eq.~\eqref{3}).
At first order in $\epsilon$, this path integral has four perturbative  contributions: The three diagrams induced by $\ca S_{1}$, $\ca S_{\alpha}$, and $\ca S_{\beta}$, as well as the change in the diffusion constant $D_{\epsilon}$. 
The simplest way of doing these calculations is to calculate with $D=1$, and finally correct for $D_{\epsilon}\neq 1$ by writing the FPT density in time of $z_t$ as
\bea\label{eq:p_stretched_g}
&& \mathbb P^{\mu,\nu}(m,t) = \mathbb G^{\mu,\nu}(m, t D_{\epsilon})
\eea
where we introduce the auxiliary probability density
\bea
&&\mathbb G^{\mu,\nu}(m, t) \\
&& = \frac{\partial}{\partial_{z_{1}}}\Big|_{z_{1}=0}
\int\limits_{z_{0}= m}^{z_t = z_1} \!\!\cD[z_t] \Theta(z_t)\,\rme^{- \ca S^0 -\ca S_{\rm d}+\varepsilon  ( \cS_1 - \alpha \cS_{\alpha} - \beta \cS_{\beta} )} \nn\\
&&\quad  +\ca O(\epsilon^{2})\ . \nn
\eea
%
%
We now use the perturbation expansion established in Ref.~\cite{WieseMajumdarRosso2010,DelormeWiese2015,DelormeWiese2016b,DelormeWiese2016}; we refer to \cite{DelormeWiese2016,DelormeThesis} for a detailed introduction, and only briefly summarise the method.
%
%

The function $\mathbb G^{\mu,\nu}(m, t)$ introduced above has the perturbative expansion 
\be
\mathbb G^{\mu,\nu} (m,t)= \rme^{{-\ca S_{\rm d}}} \big[ \mathbb G_{0} (m,t) + \epsilon \,\delta  \mathbb G   (m,t)\big] 
\ee
where 
%
\begin{align}
&\delta \mathbb G (m, t)\nn \\ \nn
&= \partial_{z_{1}}\Big|_{z_{1}=0}\int_{z_{0}= m}^{z_t = z_1} \cD[z_t] \Theta(z_t) \left(\cS_1 - \alpha\cS_{\alpha} -\beta \cS_{\beta}  \right) \rme^{ - S_0 }\\
&\stackrel!=     {\mathbb G}_1(m,t) -\alpha \mathbb G_{\alpha}(m,t) - \beta \mathbb G_{\beta}(m,t)  + \cO(\eps)\ .
\label{eq:g_sums}
\end{align}
The three auxiliary functions are defined as 
\begin{align}
\label{eq:def_g1} {\mathbb G}_1(m,t) & := \partial_{z_{1}}\int_{z_{0}= m}^{z_t = z_1} \cD[z_t] \Theta(z_t) \cS_1  \rme^{ - S_0}\Big|_{z_{1}=0}\ ,\\
\label{eq:def_galpha}  {\mathbb G}_{\alpha}(m,t) & := \partial_{z_{1}}\int_{z_{0}= m}^{z_t = z_1} \cD[z_t] \Theta(z_t) \cS_{\alpha}  \rme^{ - S_0}\Big|_{z_{1}=0} \ ,\\
\label{eq:def_gbeta}   {\mathbb G}_{\beta}(m,t) & := \partial_{z_{1}} \int_{z_{0}= m}^{z_t = z_1} \cD[z_t] \Theta(z_t) \cS_{\beta}  \rme^{ - S_0}\Big|_{z_{1}=0} \ .
\end{align}
As the term $\ca S_{\rm d}$ only depends on the initial and final point, as well as the time $T$, we were able to take it out. 
Each of the perturbations $\ca S_{1}$, $\ca S_{\alpha}$, and $\ca S_{\beta}$, defined in  Eqs.~\eq{S-alpha}-\eq{S-1} has to be evaluated   inserted into the path integral with absorbing boundaries at $z=0$.

Let us summarize the rules of this perturbative expansion, explained in detail in Ref.~\cite{DelormeWiese2016}.
The first step is to perform a Laplace transform, from the time variable $t$ to the Laplace conjugate $s$. 
This transform has two advantages: First of all, it eliminates integrals over the intermediate times. Second, the propagator \eq{16-bis}-\eq{16-cis} is exponential in the space variables, thus the latter can be integrated over.

 The next step is to   eliminate the denominator in \Eq{S-1},  using a Schwinger parametrization (Eq.~(31) of  \cite{DelormeWiese2016}), 
\begin{equation}\label{41}
\frac{1}{ {t_2-t_1}} = \int_{y>0} e^{-y (t_2-t_1)}\ .
\end{equation}
The variable $y$ on the r.h.s.\ of \Eq{41} can be interpreted as a shift in the Laplace variable $s$ associated to the time difference $t_2-t_1$, i.e.\ 
\be
s \to s+y
\ee
for all propagators between times $t_1$ and time $t_2$. For an example see the first diagram in \Eq{15} below.%

The integral over times   necessitates a cutoff $\tau$ at small times, which can be replaced by a cutoff $\Lambda$ for large $y$ (Eq.~(A3) of \cite{DelormeWiese2016}). Their relation is 
\begin{eqnarray}\nn
\lefteqn{\int_0^T \rmd t \int_0^{\Lambda} e^{-y t} \rmd y =\log (T \Lambda)+\gamma_{\rm E}+ \mathcal{O}(e^{-T \Lambda })} \\
&&\overset{!}{=} \log\!\left(\frac{T}{\tau}\right)= \int_\tau ^T  \frac{1}{t}\, \rmd t\ .~~~~~~~~~~~~~~~~~~~~~~~~~~~~~~
\end{eqnarray}
This implies the choice 
\be\label{Lambda-def}
\Lambda = \rme^{-\gamma_{\rm E}} / \tau\  .
\ee 
Finally, while the insertion of the position $x_t$ at time $t$ with $0<t<T$ leads to a factor of $x$ in the corresponding propagators, 
\be
\left <   z_t \right>_{z_0=a,z_T=b}  = \int_z P_+(a,z,t) z P_+(z,b,T-t) \ ,
\ee
the insertion of $\dot x_t$ yields a derivative   (Eq.~(A1) of  \cite{DelormeWiese2016})
\be
\left <  \dot z_t \right>_{z_0=a,z_T=b}  = 2 \int_z P_+(a,z,t) \partial_z P_+(z,b,T-t) \ .
\ee
Here $P_+(a,b,T)$ is the   Brownian  transition density introduced in Eq.~\eqref{3} in the absence of drift ($\mu=0$).

\subsection{Diagrams to be evaluated}
\label{s:Diagrams to be evaluated}
The three auxiliary functions introduced in Eqs.~\eqref{eq:def_g1}-\eqref{eq:def_gbeta} have a diagrammatic representation presented in Fig.~\ref{f:G1+G2+G3}. They give to first order in $\epsilon$   for $\mathbb G$, 
\bea \label{13}
\mathbb G^{\mu,\nu}(m,T) &:=& \exp\!\left( {-\frac{ m}2\left( \frac\mu{D_\epsilon}{+}\nu\right) -\frac T4 \left( \mu  T^{-  \epsilon} {+}\nu  T^{\epsilon}\right)^{2} } \right)\nn\\
&& \times \Big\{  {\mathbb G}_0(m,T) +  \epsilon  \Big[      {\mathbb G}_1(m,T)-   \alpha  \,  {\mathbb G}_{\alpha }(m,T)  
\w
 ~~ ~~~~~~~~~~~~~~~~~~~~~~~~~~~\, -  \beta   {\mathbb G}_{\beta }(m,T) \Big]\Big\}\ .
\eea

The zeroth order contribution ${\mathbb G}_0(m,t) $ follows from Eqs.~\eqref{32} and \eqref{p-abs},
\bea
{\mathbb G}_0(m,t) &=&\frac{m \rme^{-\frac{m^2}{4 t}}}{2 \sqrt{\pi } t^{3/2}}
\\
\tilde {\mathbb G}_{0}(m,s)&=& \rme^{{-m \sqrt s}}\ .
\eea


\subsection{Order $\epsilon$, first diagram $  {\mathbb G}_{1}$}
\label{s:G1}
The Laplace transform of the first diagram   is obtained from the insertion of $\ca S_1$ (without drift), as represented by the first diagram of figure \ref{f:G1+G2+G3}, using the Brownian propagators found in Eq.~\eqref{16-bis}. (The global factor of $2 = 2^{2}/2$ comes from a factor of $2$ for each insertion of $\dot x$, and the $1/2$ from the action.)
\begin{widetext}
\bea \label{15}
   \tilde {\mathbb G}_1(m,s) &=&  \lim_{x_{0}\to 0} \frac{ 2} {x_0} \int_0^\Lambda \rmd y\int_{x_{1}>0}\int_{{x_{2>0}}}  \tilde P_+(m,x_1,s) \partial_{x_1}  \tilde P_+(x_1,x_2,s+y)   \partial_{x_{2}} \tilde P_+(x_{2},x_0,s)   \nn \\
   &=&{ 2}\int_0^{\Lambda} \rmd y \frac{\sqrt{s} \left(\rme^{-m
   \sqrt{s}} \left(m y-2
   \sqrt{s+y}\right)+2
   \sqrt{s+y} \rme^{-m
   \sqrt{s+y}}\right)}{2 y^2} \nn\\
   & =& 
e^{ m \sqrt{s}} \left(m \sqrt{s}+1\right) \text{Ei}\left(-2 m \sqrt{s}\right)+ \rme^{-m \sqrt s}\bigg[m \sqrt{s}
   \left(\log \left(\frac{m}{2 \sqrt{s} \tau }\right)-1\right)-\log \left(2 m
   \sqrt{s}\right)-\gamma_{\rm E}
 \bigg]\ ,
\eea 
where we introduced the exponential integral function $\text{Ei}(z) = -\int_{-z}^{\infty} \rmd t \,\frac{e^{-z}}{z}$, and used \Eq{Lambda-def} to eleminate $\Lambda$. 
For the inverse Laplace transform we find using appendix C of Ref.~\cite{DelormeWiese2016b}
\checked
\bea \label{16}
{\mathbb G}_1(m,t) 
&=& {\mathbb G}_0(m,t)\Bigg[ { \mathcal{I}}\left(\frac{m}{\sqrt{2t}  }\right)+{ 2}\left(\frac{m^2}{4 t}-1\right)
   \log \left(\frac{  m^2}{\tau}\right) + 
   { \log} \left(  \frac t{\tau}\right) + \frac{(\gamma_{\rm E} -1) m^2}{2 t}-2 \gamma_{\rm E} -1\Bigg]
\ .
\eea
The special function $\mathcal{I}$ appearing in this expression was introduced in Ref.~\cite{WieseMajumdarRosso2010}, Eq.~(B53)
\begin{eqnarray}\label{defI_maintext}
\mathcal{I}(z) &=& \frac{z^4}{6}  \,_2F_2\! \left(1,1; \frac{5}{2},3; \frac{z^2}{2} \right) + \pi (1-z^2)\, \mathrm{erfi}\!\left( \frac{z}{\sqrt{2}} \right) - 3z^2 + \sqrt{2 \pi} \rme^{\frac{z^2}{2}}z +2\ ,
\end{eqnarray}
where $ \text{erfi}(z)$ is the   imaginary error function. Using the definition \eq{Lambda-def} of $\Lambda$, Eq.~(\ref{16}) 
and introducing the variable 
\be
z:=\frac{m}{\sqrt{2t}} \label{z}\ ,
\ee
${\mathbb G}_0(m,t)$ and ${\mathbb G}_1(m,t)$  can be written more compactly as 
\bea
\label{G0z}
t {\mathbb G}_0(m,t) &=& \frac{\rme^{-\frac{z^2}{2}} z}{\sqrt{2 \pi } }\ ,  \\
\label{G1z}
{\mathbb G}_1(m,t) &=&  {\mathbb G}_0(m,t)\Bigg\{ \mathcal{I} 
 (z)-\log \left(\frac{4 t z^4}{\tau }\right)+z^2 \left[\log \left(\frac{2 t z^2}{\tau
   }\right)+\gamma_{\rm E} -1\right]-2 \gamma_{\rm E} -1 \Bigg\}
\ .
\eea
Note that   there is a global prefactor of $1/t$, and  a logarithmic dependence on $t$ and $\tau$. 

\subsection{Order $\epsilon$, second diagram $  {\mathbb G}_{\alpha }$}
\label{s:Galpha}
To study perturbations with $S_\alpha$ defined in \Eq{S-alpha}, we   represent the logarithm as 
\be\label{54}
 \ln \left(\frac {t}{\tau}\right) = \int_0^\infty \frac {\rmd y}y \left[\rme^{-\tau y}-\rme^{-t y}\right] \ .
\ee
This yields for the insertion of $\ca S_\alpha$
\bea
\tilde {\mathbb G}_{{\alpha}}(m,s) &=& \lim_{x_0\to0} \frac1 {x_0} \int_0^\Lambda \frac{\rmd y}y \int_{x_{1}>0}  \Big[\tilde P_+(m,x_1,s)\rme^{{-\tau y}} -\tilde P_+(m,x_1,s+y)\Big]  \partial_{x_1}  \tilde P_+(x_1,x_0,s)  \nn\\
&=& \int_0^{\Lambda/s}  {\rmd y} \bigg[  \frac{\rme^{-m \sqrt{s}}}{\sqrt s y^2}-\frac{\rme^{-m \sqrt{s}
   \sqrt{y+1}}}{\sqrt s y^2}-\frac{m \rme^{- m \sqrt{s}-s
   \tau  y}}{2  y}\bigg] \nn\\
   &=& \frac{1}{4} m \rme^{-m \sqrt{s}} \left[2 \rme^{2 m \sqrt{s}}
   \text{Ei}\left(-2 m \sqrt{s}\right)+\log \left(\frac{4 s \tau
   ^2}{m^2}\right)+2\right] + \ca O(\Lambda^{-1})
   \label{22}
\ .
\eea
We checked that the $y$ integrand is convergent, at least as $1/y^2$ for large $y$, and has a  finite limit for $y\to 0$; thus neither $x_0$ nor $\Lambda$ are necessary as   UV cutoffs, and the $y$-integral   is finite. The $\tau$-dependence stems from the $\log(t/\tau)$ of the perturbation term. 

Doing the inverse Laplace transform using appendix C of \cite{DelormeWiese2016b}, we get with $z$ defined in Eq.~(\ref z)
\be
  \sqrt{t} {\mathbb G}_{\alpha }(m,t) =\frac{\rme^{-\frac{z^2}{2}} z^2 \left[\mathcal{I}(z)-2\right]}{2 \sqrt{\pi }
   (1-z^2)}+\frac{z\,
   \text{erfc} (\frac{z}{\sqrt{2}} )}{\sqrt{2 }
   \left(z^2-1\right)}-\frac{\rme^{-\frac{z^2}{2}} z^2 \Big[ \log\!
   \left(\frac{2 t z^2}{\tau }\right)+\gamma_{\rm E}-1\Big]}{2 \sqrt{\pi  }}  
\ ,
\ee
defining the complementary error function $\text{erfc}(z)=1-\text{erf}(z)$. Note that there is no pole at $z=1$. Indeed, for $z\to 1$  one obtains
\be
\frac{-\, _2F_2\left(1,1;\frac{5}{2},3;\frac{1}{2}\right)-4 \,
   _2F_2\left(1,1;\frac{3}{2},2;\frac{1}{2}\right)+2 \sqrt{2 e \pi }
   \left(\text{erfc}\left(\frac{1}{\sqrt{2}}\right)-3\right)+4 \pi 
   \text{erfi}\left(\frac{1}{\sqrt{2}}\right)-4 \log \left(\frac{2
   t}{\tau }\right)-4 \gamma_{\rm E} +22}{8 \sqrt{e \pi  }}
\ .
\ee

\subsection{Order $\epsilon$, third diagram $  {\mathbb G}_{\beta }$}
\label{s:Gbeta}
Using again the integral representation \eq{54}, the third diagram  for the  insertion of $\ca S_\beta$
 is read off from Fig.~\ref{f:G1+G2+G3} as 
\bea
\tilde {\mathbb G}_{\beta }(m,s) &=& \lim_{x_0\to 0} \frac1 {x_0} \int_0^\Lambda \frac{\rmd y}y \int_{x_{1}>0}   \tilde P_+(m,x_1,s)\,  \partial_{x_1} \! \Big[  \tilde P_+(x_1,x_0,s)\rme^{{-\tau y}} -  \tilde P_+(x_1,x_0,s+y) \Big] \nn\\
&=& \int_0^\infty \rmd y\left[ \frac{\sqrt{y+1} \rme^{-m \sqrt{s}}}{\sqrt s y^2}-\frac{\sqrt{y+1}
   \rme^{-m \sqrt{s} \sqrt{y+1}}}{\sqrt s y^2}-\frac{m \rme^{- m
   \sqrt{s}-s \tau  y}}{2   y}  \right] \nn\\
   &=& \frac{\rme^{-m \sqrt{s}} \left( m  \sqrt{s} \Big[2 - \log
   \left(\frac{m^2}{4 s \tau ^2}\right)\Big] +\log \left( {4 m^2
   s}\right)+2 \gamma_{\rm E} \right)}{4
   \sqrt{s}}-\frac{\rme^{m \sqrt{s}} \left(m \sqrt{s}+1\right)
   \text{Ei}\left(-2 m \sqrt{s}\right)}{2 \sqrt{s}}\ .\qquad
   \label{25}
\eea
We checked that the $y$ integrand is convergent, as it decays  at least as $1/y^{3/2}$ for large $y$, and has a finite limit for $y\to 0$, thus no UV cutoff is necessary,  and the $y$-integral   is finite. 

\end{widetext}
Doing the inverse Laplace transform using appendix C of Ref.~\cite{DelormeWiese2016b}, we get with $z$ defined in Eq.~(\ref z)
\bea
 \sqrt t\, {\mathbb G}_{\beta }(m,t) &=& \frac{\rme^{-\frac{z^2}{2}} [\mathcal{I}(z)-2]}{2 \sqrt{\pi  }
   \left(1-z^2\right)}+\frac{z\,
   \text{erfc} (\frac{z}{\sqrt{2}} )}{\sqrt{2 }
   \left(z^2-1\right)}
\w
   +\frac{\rme^{-\frac{z^2}{2}} z^2 \left[1-\log
    (\frac{t}{\tau } )\right]}{2 \sqrt{\pi }} \ .
\eea

\subsection{Combinations}
\label{s:Combinations}
Let us remind that  in the drift-free case the result for ${\mathbb G}_0(z)$ is given in \Eq{G0z}, while
${\mathbb G}_1(z)$ is given in Eq.~\eq{G1z}.
Let us now turn to the corrections for drift. 
While ${\mathbb G}_{\alpha}$ and ${\mathbb G}_{\beta }$ are the appropriate functions
for the calculations, we finally need the corrections for linear
drift $\mu$ and non-linear drift $\nu$. Demanding that
\be
\alpha {\mathbb G}_{\alpha} + \beta {\mathbb G}_{\beta } \stackrel!=  \mu {\mathbb G}_{\mu}+\nu {\mathbb G}_{\nu}\ ,
\ee
and using Eqs.~\eq{14a} and \eq{14b} yields
\bea
\sqrt t {\mathbb G}_{\mu}(m,t)&=& { \sqrt t \big[{\mathbb G}_{\alpha }(m,t )+
{\mathbb G}_{\beta }(m,t )\big] } \nn\\ \nn
&=& -\frac{\rme^{-\frac{z^2}{2}} \left(z^2{+}1\right) [\mathcal{I}(z){-}2]}{{ 2}
   \sqrt{\pi  } \left(z^2{-}1\right)}+\frac{ { \sqrt{2} }\, z\,
   \text{erfc} (\frac{z}{\sqrt{2}} )}{
    z^2{-}1}
\\
&&-\frac{\rme^{-\frac{z^2}{2}} z^2 \left[\log \!
   \left(\frac{2 t^2 z^2}{\tau ^2}\right)+\gamma_{\rm E} -2\right]}{{ 2}
   \sqrt{\pi  }} 
\eea
\bea
 \sqrt t  {\mathbb G}_{\nu}(m,t )&=&\sqrt t{\big[   {\mathbb G}_{\beta }(m,t )-
{\mathbb G}_{\alpha }(m,t) \big]} \nn\\&=& \frac{\rme^{-\frac{z^2}{2}} [\mathcal{I}(z){-}2]}{{ 2} \sqrt{\pi  }}+\frac{\rme^{-\frac{z^2}{2}} z^2 \left[\log  (2
   z^2 ){+}\gamma_{\rm E} \right]}{{ 2} \sqrt{\pi }}\ . \w
\eea
The perturbative contributions  can be grouped together as, cf.~Eqs.~\eqref{eq:g_sums} and \eqref{13}
\bea
\lefteqn{{\mathbb G}(m,t) := \exp\!\left( {-\frac{ m}2\left[ \frac\mu{D_\epsilon}+\nu\right] -  \frac t4 \left[\mu  t^{-     \epsilon} +\nu  t^{ \epsilon}\right]^{2} } \right)}\nn\\
&& \!\times \left\{  {\mathbb G}_0(m,t) + { \epsilon } \Big[    \, {\mathbb G}_1(m,t){-}  \mu  \,  {\mathbb G}_{\mu}(m,t) {-} \nu \,  {\mathbb G}_{\nu}(m,t) \Big]\right\}.\w
\eea
This expression is to this order equivalent to 
 \bea\label{80}
&&\!\!\!\! {\mathbb G}(m,t) := \exp\!\left( {-\frac{ m}2\left[ \frac\mu{D_\epsilon}+\nu\right] -  \frac t4 \left[\mu  t^{-    \epsilon} +\nu  t^{ \epsilon}\right]^{2} } \right) \w \times {\mathbb G}_0(m,t)\nn\\
&& \times \exp \left(  { \epsilon}  \frac{     \, {\mathbb G}_1(m,t)-  \mu  \,  {\mathbb G}_{\mu}(m,t) - \nu\,   {\mathbb G}_{\nu}(m,t)  }{{\mathbb G}_0(m,t)}\right)\ .
\eea
See \cite{Wiese2018}, Sec. IV.C for a discussion of why it is better to write the perturbative corrections in an exponential form.
\subsection{Scaling and corrections from the diffusion constant, final result}
\label{s:Scaling and corrections from the diffusion constant, final result}
The natural scaling variable for fBm is not $z$, but 
\be
y:=\frac{m}{\sqrt 2 t^{H}}\ .
\ee
This will induce some corrections (cf.~Eq.\eqref{eq:p_stretched_g}). 
Consider  
\be
\frac{\rme^{-\frac{y^2}{2}} y}{\sqrt{2 \pi } } = \frac{\rme^{-\frac{z^2}{2}} z}{\sqrt{2
   \pi }} \Big[ 1 + 
    (z^2-1 ) \epsilon  \log
   (t)\Big]+\ca O(\epsilon ^2)\ .
\ee
There is also a correction to the diffusion constant, 
\be
D_{\epsilon} \simeq (e\tau)^{{2\epsilon}} \ .
\ee 
According to \Eq{eq:p_stretched_g}, this implies that 
\bea
\lefteqn{ \mathbb P (m,t) = {\mathbb G}(m,t D_\epsilon)}
 \nn\\
 &=& \frac{\rme^{-\frac{y^2}{2}} y}{\sqrt{2 \pi }t D_{\epsilon} }  
 \times \exp\!\left( -\frac{ m}2\left[ \frac\mu{D_\epsilon}+\nu\right] \right) 
\w
\times \exp\!\left( 
 -\frac {D_{\epsilon}  t}4 \Big[ \mu^2 (D_{\epsilon}  t)^{-    2\epsilon} +\nu^2 (D_{\epsilon} t)^{2\epsilon}\Big]  \right)  \nn\\
&& \times \exp \!\bigg(  \epsilon \bigg[  \frac{     {\mathbb G}_1(m,t)-   \mu  \,  {\mathbb G}_{\mu  }(m,t) - \nu  \, {\mathbb G}_{\nu}(m,t)  }{{\mathbb G}_0(m,t)} \w
~~~~~~~~~~~~~~~~-     (y^2-1 )   \log
   (t)\bigg]\bigg)
\eea   
Note that we used the factored form \eq{80} to make appear the ratios of    $\mathbb G_1$, $\mathbb G_\mu$ and $\mathbb G_\nu$ with $\mathbb G_0$, yielding (relatively simple) special functions $\ca F_1$, $\ca F_\mu$  and $\ca F_\nu$ defined below. Regrouping terms yields
\bea  
\lefteqn{{\mathbb P}(m,t) =\frac{\rme^{-\frac{y^2}{2}} y^{\frac1H-1}}{\sqrt{2 \pi }t } }
\w   
    \times \exp\!\left( -{\frac {\mu m^{1-2\epsilon/H}  }2 y^{{2\epsilon}}  -\frac{ \nu m}2   y^\epsilon - \frac t4 \left[\mu  t^{-     \epsilon} +\nu  t^{ \epsilon}\right]^{2} } \right) 
\nn\\
&&    
    \times \exp \!\left(  \epsilon \Big[\ca F_{1}(y) - \mu m \ca F_{\mu}(y) -\nu m \ca F_{\nu}(y)\Big]\right)
\ .
\label{eq:p_in_terms_of_f}
\eea
To order $\epsilon$, this can be  rewritten in a more intuitive form as 
\bea   \label{36}
\lefteqn{t\, \mathbb P(m,t) = \frac{y^{\frac1H-1}}{\sqrt{2 \pi } }\times}
\\&&
 \times \exp\!\Bigg(\! {-}\frac {y^2}{2} + \epsilon \Big[\ca F_{1}(y){+}\ca F_1^0\Big]-{\mu m^{\frac 1 H-1}  } y^{ 2 \epsilon}  \left[\frac 12{+}\epsilon \ca F_{\mu}(y) \right] 
\w
~~~~~~ -  \nu m y^{ 2\epsilon}\left[ \frac 12{+}\epsilon \ca F_\nu(y)\right]  
  - \!\frac{m^{2}}{8 y^{2}}  \!\left [ 
 \mu\! \left( \frac {2 y^2}{m^2}\right)^{\!\!\frac \epsilon H}  \!\!\!\! +\nu 
 \right]^{2} \Bigg).\nn
\eea
Note that since our  expansion is restricted to the first order in $\epsilon$, in expressions like 
\be
\frac{1}H-1 = 1- 4 \epsilon  + \ca O(\epsilon^2)\ , \qquad 1-\frac1{2H} = 2\epsilon+ \ca O(\epsilon^2)\ ,
\ee
we have no means to distinguish between left- and right-hand side. Some choices are given by scaling, as the prefactor of $y^{\frac1H-1}$, or seem natural, others are educated guesses. 

Finally, we wish to rewrite 
\Eq{36} (a density in time) as a density in  $y$, given distance $m$ from the absorbing boundary for the  starting  point.
Using that  
\be\label{measure-change}
\frac {\rmd t}t = \frac 1H \frac{\rmd y} y\ ,
\ee
this yields \be\label{74}
\ca P(y|m,\mu,\nu) = \ca P_{>}(y|m,\mu,\nu) + \mathbf P_{\rm escape}(m,\mu,\nu) \delta(y)\ .
\ee
The function $\ca P_{>}(y|m,\mu,\nu) $ is  equivalent to \Eq{36} after the change in measure \eq{measure-change}, 
\bea\label{P(y|m)-final}
\lee{\ca P_{>}(y|m,\mu,\nu) = \frac{y^{  \frac1H-2}}{\sqrt{2 \pi } H}  \times}
\\&&
\!\times\! \exp\!\Bigg( \!{-}\frac {y^2}{2} + \epsilon\Big[ \ca F_{1}(y) {+}\ca F_1^0\Big]
 -{\mu m^{\frac 1 H-1}  } y^{  2 \epsilon}  \left[\frac 12{+}\epsilon F_{\mu}(y) \right]
\w 
~~~~~~~~~~~~~ -  \nu m y^{2 \epsilon}\left[ \frac 12{+}\epsilon \ca F_\nu(y)\right]  
  - \!\frac{m^{2}}{8 y^{2}}  \!\left [ 
 \mu\! \left( \frac {2 y^2}{m^2}\right)^{\!\!\frac \epsilon H}  \!\!\!\! +\nu 
 \right]^{2} \Bigg). \nn
\eea
Some trajectories escape, which we count as absorption time $t=\infty$, equivalent to $y=0$,   resulting into the contribution proportional to  $\delta(y)$ in \Eq{74}, with amplitude 
\be
 \mathbf P_{{\rm escape}}(m,\mu,\nu ) = 1- \mathbf P_{\rm abs}(m,\mu,\nu) \ ,
\ee
where 
\be
\mathbf P_{\rm abs}(m,\mu,\nu): = \int_{0}^\infty \rmd y\,  \ca P_>(y|m,\mu,\nu) \ .
\ee It is evaluated in  the next section, see Eqs.~\eq{P-abs-final1}-\eq{P-abs-final3}.

The three special functions appearing in Eq.~\eqref{eq:p_in_terms_of_f} are defined as follows:
First, the drift-free contribution are
\bea\label{F1-def}
\lefteqn{   {\ca F}_{1}(y) +{\ca F}_{1}^0 }\nn\\
   &:=&   \frac{\mathbb G_{1}(y)}{\mathbb G_{0}(y)} -      ({ y^2-1} )  \Big[ \log
   (t/\tau) -1\Big] + 4\ln y  \\
&=&\mathcal{I}(y)+y^2 \left(\log \left(2 y^2\right){+}\gamma_{\rm E}\right)-2 \left( 
   \gamma_{\rm E}+1+\ln 2 \right)\nn
\ .
\eea
The   conventions are  s.t.\ ${\ca F}_{1}(y)$ agrees with Refs.~\cite{WieseMajumdarRosso2010,DelormeWiese2016,DelormeWiese2015}, i.e.\ ${\ca F}_{1}(0)=0$. The constant part ${\ca F}_{1}^0$ is equivalent to a change in normalization, $\ca N = \exp( {-\epsilon\ca F_1^0})$, which for the drift-free case was of no interest  \cite{WieseMajumdarRosso2010,DelormeWiese2016,DelormeWiese2015}, as there the absorption probability is one, which is not the case with drift. 
In the chose convention,  \checked
\bea\label{F1}
\ca F_{1}(y) 
&=&\mathcal{I}(y)+ y^2 \left(\log \left(2 y^2\right)+\gamma_{\rm E}\right)-2 \ ,\ \\
\ca F_1(0) &=&0 \ ,\\
{\ca F}_{1}^0&=& -2 \left( \gamma_{\rm E}+\ln 2\right) \ .
\eea
Its   asymptotic expansions for small and large $y$ are
\bea
\ca F_{1}(y)&=& 2 \sqrt{2 \pi } y+y^2 \left(\log \left(2 y^2\right)+\gamma_{{\rm E}}
   -3\right)-\frac{1}{3} \sqrt{2 \pi }
   y^3\nn\\
   && +\frac{y^4}{6}  -\frac{1}{30} \sqrt{\frac{\pi }{2}}
   y^5+\frac{y^6}{90}-\frac{1}{420} \sqrt{\frac{\pi }{2}}
   y^7+\frac{y^8}{1260}\nn\\
   && -\frac{\sqrt{\frac{\pi }{2}}
   y^9}{6048}+\frac{y^{10}}{18900}  +\ca O(y^{11}) \ ,\\
\ca F_{1}(y)&=&   \log (y^{2}/2)+1-\psi \left(\textstyle \frac{1}{2}\right) +\frac{1}{2 y^2}-\frac{1}{2
   y^4}+\frac{5}{4 y^6} \nn\\
   && -\frac{21}{4 y^8}+\frac{63}{2
   y^{10}}+\ca O(y^{{-11}})\ .
   \label{F1-large-y}
\eea\begin{figure*}
\fig{5.7cm}{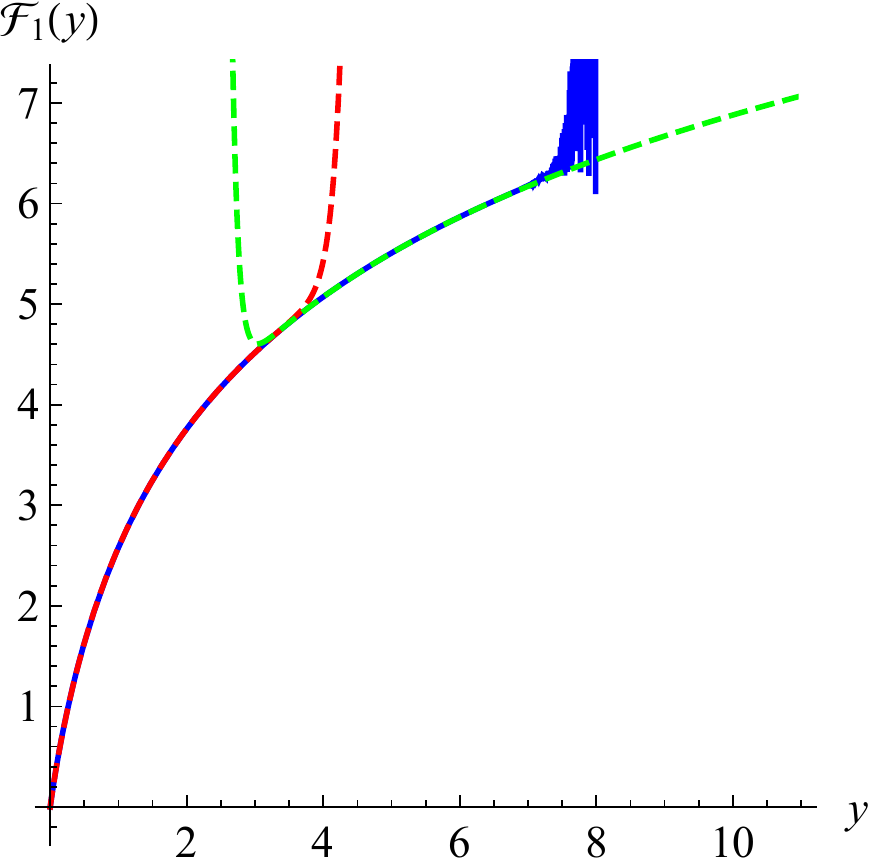}\hfill\fig{5.7cm}{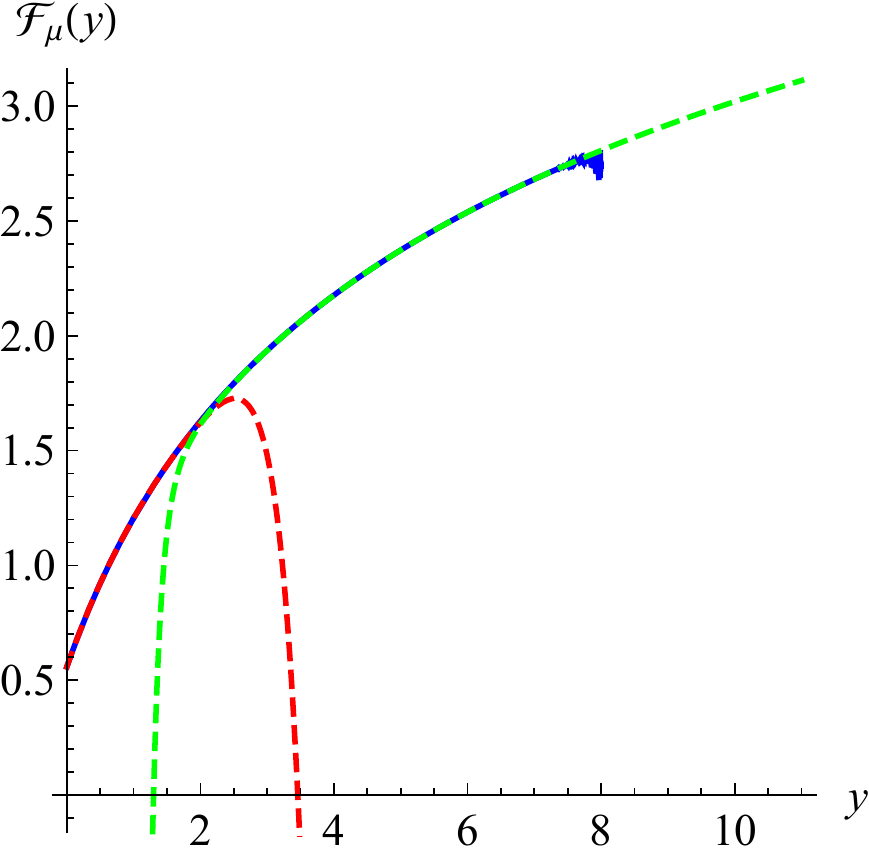}\hfill\fig{5.8cm}{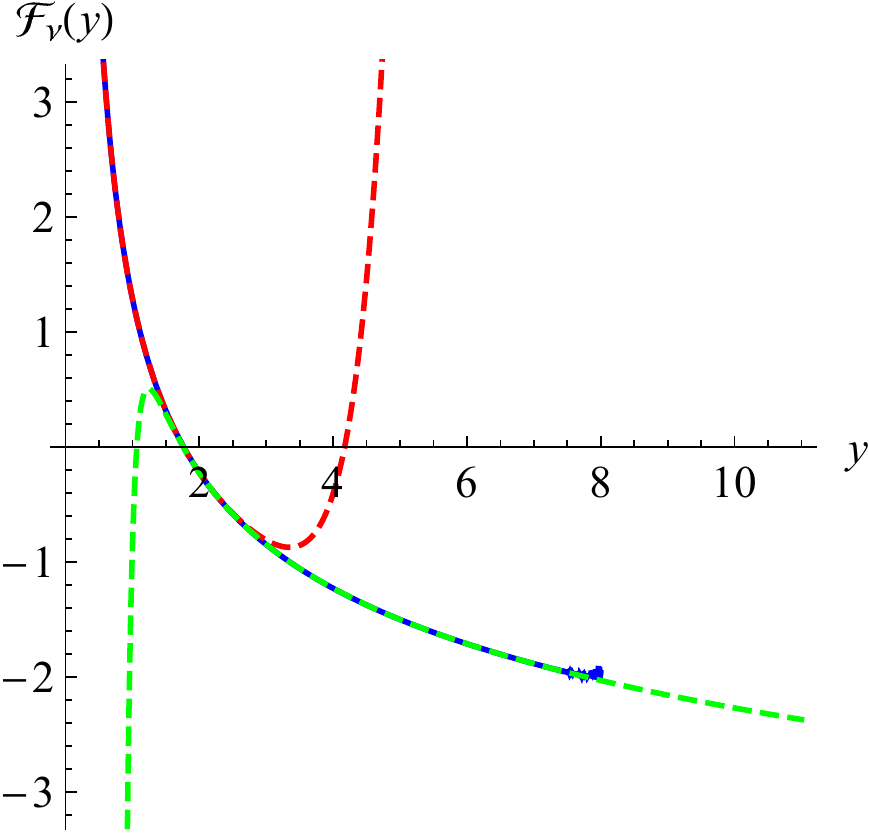}
\caption{Left: The function $\ca F_{1}(y)$ (blue, solid), with its asymptotic expansions (red and green dashed). Middle: {\it ibid.} for $\ca F_\mu(y)$. Right: {\it ibid.} for $\ca F_\nu(y)$. Numerical measurements are presented on Figs.~\ref{f:F1-num}, \ref{f:Fmu-num} and \ref{f:Fnu-num}.
}
\end{figure*}
\Eq{F1} is  equivalent to Eqs.~(55) in \cite{WieseMajumdarRosso2010}, and (56) in \cite{DelormeWiese2016}. 

The second function is for the drift proportional to $\mu$, 
\be\label{Fmu-def}
{\ca F}_{\mu}(y) := \frac{ {\mathbb G}_{\mu}(m,t)}{m \mathbb G_{0}(m,t)}+   \partial_{\epsilon}\bigg|_{\epsilon=0}\left({ \frac{m^{4 \epsilon}}{2 D_{\epsilon} y^{2\epsilon}}}\right)\ .
\ee
It is evaluated as 
\bea\label{Fmu}
{\ca F}_{\mu}(y) 
& =& \frac{  \left(y^2+1\right) [\mathcal{I}(y)-2]}{2 y^{2} (1-y^2)}+ \frac{\sqrt{2 \pi } \rme^{\frac{y^2}{2}}
   \text{erfc}\left(\frac{y}{\sqrt{2}}\right)}{y
   \left(y^2-1\right)} \nn\\
   && +\frac{1}{2} \Big[\log (2)-\gamma_{\rm E} \Big]\ .   
\eea
Its asymptotic expansions are 
\bea   
{\ca F}_{\mu}(y) &=&    \frac{1}{2} \Big[1-\gamma_{{\rm E}} +\log (2)\Big]+\frac{1}{3} \sqrt{2 \pi }
   y-\frac{y^2}{4}+\frac{1}{15} \sqrt{\frac{\pi }{2}}
   y^3 \nn\\
   && -\frac{y^4}{36}+\frac{1}{140} \sqrt{\frac{\pi }{2}}
   y^5-\frac{y^6}{360}+\frac{\sqrt{\frac{\pi }{2}}
   y^7}{1512}-\frac{y^8}{4200} 
   \w
   +\frac{\sqrt{\frac{\pi }{2}}
   y^9}{19008}-\frac{y^{10}}{56700}  + \ca O(y^{11})\ ,\\
{\ca F}_{\mu}(y) &=&   \log
   (2 y) +   \frac{  \log (2y^{2}) +\gamma_{{\rm E}} -1}{2 y^2}+\frac{3}{4
   y^4}-\frac{5}{4 y^6}+\frac{35}{8 y^8} 
\w
   -\frac{189}{8
   y^{10}} +\ca O(y^{{-11}})\ .
\eea
Note that we added some { strangely looking} factors into the result (\ref{P(y|m)-final}). The factor $m \times m^{-\frac{2\epsilon}{H} }=m^{\frac1 H-1}$ accounts for the dimension of the diffusion constant, $m/D_\epsilon \sim m \tau^{-2\epsilon}$, and takes out the term $\ln(m)$  from $\ca F_{\mu}(y)$. We  moved out also a remaining term $\sim \ln y$.

The third function is for the drift proportional to $\nu$, 
\be \label{Fnu-def}
\ca F_{\nu}(y) := \frac{\mathbb {\mathbb G}_{\nu}(y)}{{\mathbb G}_0(y) m}  - \ln (y)\ .
\ee
It is evaluated as 
\be \label{Fnu}
\ca F_{\nu}(y)
   = \frac{\mathcal{I}(y)-2}{2 y^2 {  }}+\frac{ \log  (2 )+\gamma_{\rm E} }{2}\ .
\ee
Its asymptotic expansions read
\bea   \label{Fnu-samll}
  \ca F_{\nu}(y)&=&  \frac{\sqrt{2 \pi }}{y}+\frac{-3+\gamma_{\rm E} +\log
   (2)}{2}-\frac{1}{3} \sqrt{\frac{\pi }{2}}
   y+\frac{y^2}{12}
\w
   -\frac{1}{60} \sqrt{\frac{\pi }{2}}
   y^3+\frac{y^4}{180}-\frac{1}{840} \sqrt{\frac{\pi }{2}}
   y^5+\frac{y^6}{2520}
\w
-\frac{\sqrt{\frac{\pi }{2}}
   y^7}{12096}+\frac{y^8}{37800}-\frac{\sqrt{\frac{\pi }{2}}
   y^9}{190080}+\frac{y^{10}}{623700}\w
   +\ca O(y^{11}) \\
   \ca F_{\nu}(y)&=&-\log (y)+\frac{2 \log (y)+\gamma_{\rm E} +1+\log (2)}{2 y^2}+\frac{1}{4
   y^4}-\frac{1}{4 y^6}
\w   
   +\frac{5}{8 y^8}-\frac{21}{8
   y^{10}}
+\ca O(y^{-11}) \label{Fnu-large}
\eea
Using \Eq{P(y|m)-final} for small $y$, there is a problem when $\epsilon \nu <0$, since then the combination (second-to-last term in the exponential)
\be
- \epsilon \nu m y^{2 \epsilon}\left[ \frac 12{+}\epsilon \ca F_\nu(y)\right]  \stackrel{y\to 0 }\longrightarrow - \epsilon \nu m  \sqrt{2\pi} y ^{2\epsilon-1} \approx - 2 \nu \sqrt \pi t^{H}.
\ee
diverges (at least for $\frac14<H<\frac12$), which is amplified since it appears inside the exponential. We propose to use the following Pad\'e variant, which seems to work well numerically, 
\be\label{Pade}
\left[ \frac 12{+}\epsilon \ca F_\nu(y)\right]  \stackrel{\epsilon<0,\, \nu>0}{-\!\!\!-\!\!\!-\!\!\!-\!\!\!-\!\!\!-\!\!\!\longrightarrow}  \frac 1{2-4 \epsilon \ca F_\nu(y)}\ .
\ee
While $\ca F_\nu (y)$ diverges for small $y$, this  is at leading order nothing but a normalization factor depending on $\nu t^{H}$. 

All three functions $\ca F_{1}(y)$,  $\ca F_{\mu}(y)$  and  $\ca F_{\nu}(y)$ are measured in section \ref{s:Numerics}, see figures \ref{f:F1-num}, \ref{f:Fmu-num}, and \ref{f:Fnu-num}.

\subsection{Absorption probability}
\label{s:Absorption probability}
From \Eq{13}, we obtain, $  \mathbf P_{\rm abs}(m,\alpha,\beta)$
\begin{align}\label{86}
\lefteqn{ {\mathbf P}_{\rm abs} (m,\alpha,\beta)= \int_0^\infty\! \rmd t \, \mathbb G(m,t  D_{\epsilon}) }\nn\\
&= \int\limits_0^\infty \rmd t \,\exp\!\left(\! {-}\frac{ m}2\!\left[ \frac\mu{D_\epsilon}{+}\nu\right]\! {-}  \frac t4 \left[\mu  t^{-     \epsilon} {+}\nu  t^{ \epsilon}\right]^{2}  \right)  {\mathbb G}_0(m,t  D_{\epsilon})\nn\\
&+ { \epsilon} \int\limits_0^\infty \rmd t  \exp\!\left( {-\frac{ m}2 \beta - \frac t4 \beta^{2} } \right) \times
\wa
\qquad \times \Big[   {\mathbb G}_1(m,t)-   \alpha  \,  {\mathbb G}_{\alpha }(m,t) - \beta  {\mathbb G}_{\beta }(m,t) \Big]+ \ca O(\epsilon ^2) \nn\\
&= \exp\!\left( -\frac{ m}2 \left[ \frac\mu{D_\epsilon}+\nu\right]\right)
\wa
\times \Bigg\{ \int\limits_0^\infty \rmd t \,\exp\!\left(  -  \frac t4 \left[\mu  t^{-     \epsilon} +\nu  t^{ \epsilon}\right]^{2}  \right)  {\mathbb G}_0(m,t  D_{\epsilon})\nn\\
& \qquad +{ \epsilon}   \Big[   \tilde {\mathbb G}_1(m,s)-   \alpha  \,  \tilde {\mathbb G}_{\alpha }(m,s) -  \beta  \tilde {\mathbb G}_{\beta }(m,s) \Big] \Big|_{\sqrt s = |\beta|/2} \Bigg\}
\wa + \ca O(\epsilon ^2)
\ .
\end{align}
Here $\tilde {\mathbb G}_1(m,s)$ is given by \Eq{15}, $\tilde {\mathbb G}_{\alpha}(m,s)$ by \Eq{22}, and $\tilde {\mathbb G}_{\beta }(m,s)$ by \Eq{25}. 
We still need the integral
\checked
\bea
\lee{ \int_0^\infty \rmd t \,\exp\!\left(  -  \frac t4 \left[\mu  t^{-     \epsilon} +\nu  t^{ \epsilon}\right]^{2} \right) {\mathbb G}_0(m,t D_{\epsilon})}\nn\\
&&\qquad = \rme^{-|\beta|m/(2 \sqrt D_{\epsilon}) }+ \frac{\alpha  \beta}2 \epsilon\, {\mathbb G}_{3}(m,\beta) \ ,\\
&& {\mathbb G}_{3}(m,\beta ) =
\int_0^\infty \rmd t \,    \rme^{-\frac{\beta ^2 t}{4}} t \log (t) {\mathbb G}_0(m,t)\ .~~~~~~~
\eea
The last expression can be calculated as  \checked
\bea
\lefteqn{   {\mathbb G}_{3}(m,\beta):=\int_0^\infty \rmd t \,    \rme^{-\frac{\beta ^2 t}{4}} t \log (t) {\mathbb G}_0(m,t)} \nn\\
&=& \partial_\kappa \Big|_{\kappa=0}\int_0^\infty \rmd t \,    \rme^{-\frac{\beta ^2 t}{4}} t^{1+\kappa} {\mathbb G}_0(m,t)  \nn\\
&=& \partial_\kappa \Big|_{\kappa=0}  \frac{|\beta| ^{-\kappa -\frac{1}{2}} m^{\kappa +\frac{3}{2}} K_{\kappa
   -\frac{1}{2}}\!\left(\frac{m |\beta| }{2}\right)}{\sqrt{\pi }} 
\nn\\ 
&=&  -\frac{m^{3/2}  \partial_{\kappa}\Big|_{\kappa=0}{K}_{\kappa-\frac12 }\!\left(\frac{|\beta| 
   m}{2}\right)}{\sqrt{\pi  |\beta| }}+\frac{m \rme^{-\frac{|\beta|  m}{2}} \log
   \left(\frac{m}{|\beta| }\right)}{|\beta| }\nn\\
&=&-\frac{m e^{\frac{m \left|
\beta \right| }{2}} \text{Ei}(-m
   \left| \beta \right| )}{\left| \beta \right| } +\frac{m e^{-\frac{m \left| \beta \right| }{2}} \log \left(\frac{m}{\left| \beta \right|
   }\right)}{\left| \beta \right| }
\nn\\
   &=& \frac{m }{|\beta| }[-2 \log (|\beta| )-\gamma_{\rm E} ]+\frac{1}{2} m^2 \big[ -2 \log (m)-\gamma_{\rm E}
   +2)\big] 
\w
+ \ca O(m^3)\ ,
\eea
where ${K}_n(z)$ denotes the modified Bessel function of the second kind. With the above formulas, \Eq{86} is rewritten as \pagebreak[6]
\bea
\lefteqn{  \mathbf P_{\rm abs}( m, \alpha,  \beta) =  \rme^{- {m (\beta+|\beta|)}/2}  \bigg\{ 1 +\epsilon  \,\rme^{|\beta|m/2} \times}
\nn \\
&& \times  \bigg[ 
\frac{\alpha \beta}2    {\mathbb G}_{3} (m,\beta) + \frac{\alpha {+}\beta{+}|\beta|}{2} m  (1{+}\log \tau )\rme^{-|\beta|m/2}
\w
~~~~~~ +      \tilde {\mathbb G}_1(m,s) -   \alpha    \tilde {\mathbb G}_{\alpha }(m,s) - \beta  \tilde {\mathbb G}_{\beta }(m,s)  \bigg]_{\sqrt s =
\frac{|\beta|}2}\nn\\
&& + \ca O(\epsilon^2)
\bigg\}
\ .
\eea
We note the exact relations, which can   be verified numerically, 
\begin{align}
\label{cancel1}
& \tilde {\mathbb G}_1(m,s)+2\sqrt{s} \, \tilde {\mathbb G}_{\beta }(m,s) = 0\ ,
\\
\label{cancel2}
&  {\mathbb G}_{3}
(m,\beta)|\beta|+2\tilde {\mathbb G}_\alpha (m,s) \nn\\
&\qquad\qquad ~~~~~ -m (1+ \log  \tau ) \rm e^{-\frac{m \left| \beta \right| }{2}}
  \Big|_{\sqrt s 
= \frac{|\beta|}2}=0 
\ .
\end{align}
Let us   analyse $\mathbf P_{\rm abs}$ separately for $\beta<0$ and $\beta>0$, starting with the former. Using both cancelations in Eqs.~\eq{cancel1} and \eq{cancel2}, we find 
\be
  \mathbf P_{\rm abs}(  \alpha,  \beta<0) =1 + \ca O(\epsilon^2)\ .
\ee
Thus there is no change in normalisation for a drift towards the absorbing boundary. 
For $\beta>0$, we find again with the use of Eqs.~\eq{cancel1} and \eq{cancel2}
\bea \label{68}
\lefteqn{  \mathbf P_{\rm abs}(  \alpha,  \beta>0) =  \rme^{- m  \beta} \times}
 \\
&\times& \!\!\bigg\{ 1+\epsilon \,\bigg[  
(\alpha {+} \beta ) m  (1+\log \tau ) 
\w
+    2   \rme^{ \beta m/2} \Big( \tilde {\mathbb G}_1(m,s) -    \alpha    \tilde {\mathbb G}_{\alpha }(m,s)    \Big)_{\sqrt s =
\frac{\beta}2} \bigg] + \ca O(\epsilon^2)
\bigg\}.\nn
\eea
For what follows, we note regularity of the combination
$
\text{Ei}(-x)-\log (x)-\gamma_{{\rm E}} 
$.
We can write \Eq{68} as 
\bea \label{70}
\lefteqn{ \mathbf P_{\rm abs}(m,  \alpha,  \beta) =  \rme^{- m  \beta} \times}
\nn \\
&\times& \!\!\bigg\{ 1+\epsilon \,\bigg[ (m (\beta -\alpha )+2) \left(e^{\beta  m} \text{Ei}(-m \beta )-\log (\beta  m)-\gamma_{\rm E}
   \right) \nn\\
   && ~~~~~~~~~~~~~~~ -\alpha  m (2 \log (\beta )+\gamma _{\rm E})+\beta  m (2 \log (m)+\gamma_{\rm E} ) \bigg] 
\w
~~~+ \ca O(\epsilon^2) \bigg\} \nn\\
\lefteqn{ =  \rme^{- m  \beta} \times}
\nn \\
&\times& \!\!\bigg\{ 1+\epsilon \,m\bigg[  2 (\beta {-}\alpha ) \log (\beta )-\gamma_{\rm E} ( \alpha {+}3   \beta)
-2 \beta +4 \beta  \log (m)  \bigg] 
\w
~~~+ \ca O(\epsilon^2) + \ca O(m^{2}\epsilon) \bigg\}
\ .
\eea
As the asymptotic expansion in the last line shows, a common resummation is possible; passing to variables $\mu$ and $\nu$, it reads
\bea\label{97}
\mathbf P_{\rm abs}(m,\mu,\nu) &=& \exp\!\Big(\! -m^{\frac1H-1} \mu \, \big[1+  2 (1-\gamma_{\rm E} ) \epsilon \big] 
\w
~~~~ -m ^{\frac1H-1} \nu (\mu{+}\nu)^{\frac1H-2}  \, \big[1{+}  2 (1{-}2\gamma_{\rm E} ) \epsilon \big]  \Big)\!\!
\w  + \ca O(\epsilon^2) + \ca O(m^{2}\epsilon) 
\ .
\eea
\smallskip

\noindent This formula represents the leading behavior of $\mathbf P_{\rm abs}(m,\mu,\nu) $ for small $m$; thus terms of order   $\cO(m^2\eps)$ could be neglected. 
Note that the (inverse) powers of $H$ were chosen s.t.\ the resulting object is scale invariant. Expanding in $\epsilon$ leads back to \Eq{70}. 
One finally arrives at \begin{widetext}
\bea\label{P-abs-final1}
\mathbf P_{\rm abs}(m,\mu,\nu) &=& \exp\!\Bigg(\! -m^{\frac1H-1} \bigg\{ \mu \, \Big[1+  2 (1-\gamma_{\rm E} ) \epsilon \Big] 
+ \nu (\mu+\nu)^{\frac1H-2}  \, \Big[1+  2 (1-2\gamma_{\rm E} ) \epsilon \Big] \bigg\}
\nn\\
& &~~~~~~~~\, +\epsilon \bigg\{ 2 (m \nu +1) \Big[e^{m (\mu +\nu )} \text{Ei}\big(-m (\mu +\nu )\big)-\log \big(m (\mu +\nu )\big)-\gamma_{\rm E}
   \Big]
\w
~~~~~~~~~~~~~~~~~-2 m (\mu +\nu ) \Big[\log \big(m (\mu +\nu )\big)+\gamma_{\rm E} -1\Big]
   \bigg\} \Bigg) + \ca O(\epsilon^2)\ .
\eea
In order that this formula be invariant under  $m \to \lambda m$, $\mu\to \lambda^{1-\frac1H} \mu $ and $\nu \to \lambda^{-1} \nu$, we can either replace $m \mu $ by  $m \mu^{\frac{H}{1-H}}$, or   $m^{\frac1H-1} \mu$. The first version is
\bea\label{P-abs-final2}
\mathbf P_{\rm abs}^{(a)}(m,\mu,\nu) &=& \exp\!\Bigg(\! -m^{\frac1H-1} \bigg\{ \mu \, \Big[1+  2 (1-\gamma_{\rm E} ) \epsilon \Big] 
+ \nu \Big(\mu^{\frac{H}{1-H}}+\nu\Big)^{\frac1H-2}  \, \Big[1+  2 (1-2\gamma_{\rm E} ) \epsilon \Big] \bigg\}
\nn\\
& &~~~~~~~~\, +\epsilon \bigg\{ 2 (m \nu +1) \Big[e^{m \big(\mu^{\frac{H}{1-H}} +\nu \big)} \text{Ei}\Big(-m \big(\mu^{\frac{H}{1-H}} +\nu \big)\Big)-\log\! \Big(m \big(\mu^{\frac{H}{1-H}} +\nu \big)\Big)-\gamma_{\rm E}
   \Big]
\w
~~~~~~~~~~~~~~~~~-2 m \Big(\mu^{\frac{H}{1-H}} +\nu \Big) \Big[\log\! \Big(m\big (\mu^{\frac{H}{1-H}} +\nu\big )\Big)+\gamma_{\rm E} -1\Big]
   \bigg\} \Bigg) + \ca O(\epsilon^2)\ .
\eea
The alternative second version is 
\bea\label{P-abs-final3}
\mathbf P_{\rm abs}^{(b)}(m,\mu,\nu) &=& \exp\!\Bigg(\! -m^{\frac1H-1} \bigg\{ \mu \, \Big[1+  2 (1-\gamma_{\rm E} ) \epsilon \Big] 
+ \nu (\mu^{\frac{H}{1-H}}+\nu)^{\frac1H-2}  \, \Big[1+  2 (1-2\gamma_{\rm E} ) \epsilon \Big] \bigg\}
\nn\\
& &~~~~~~~~\, +\epsilon \bigg\{ 2 (m \nu +1) \Big[e^{m^{\frac1H-1} \mu+m \nu} \text{Ei}\big(-m^{\frac1H-1} \mu-m \nu\big)-\log \big(m^{\frac1H-1} \mu+m \nu\big)-\gamma_{\rm E}
   \Big]
\w
~~~~~~~~~~~~~~~~~-\Big( m^{\frac1H-1} \mu+m \nu\Big) \Big[\log \big(m^{\frac1H-1} \mu+m \nu\big)+\gamma_{\rm E} -1\Big]
   \bigg\} \Bigg) + \ca O(\epsilon^2)\ .
\eea
\begin{figure*}
\fig{8.8cm}{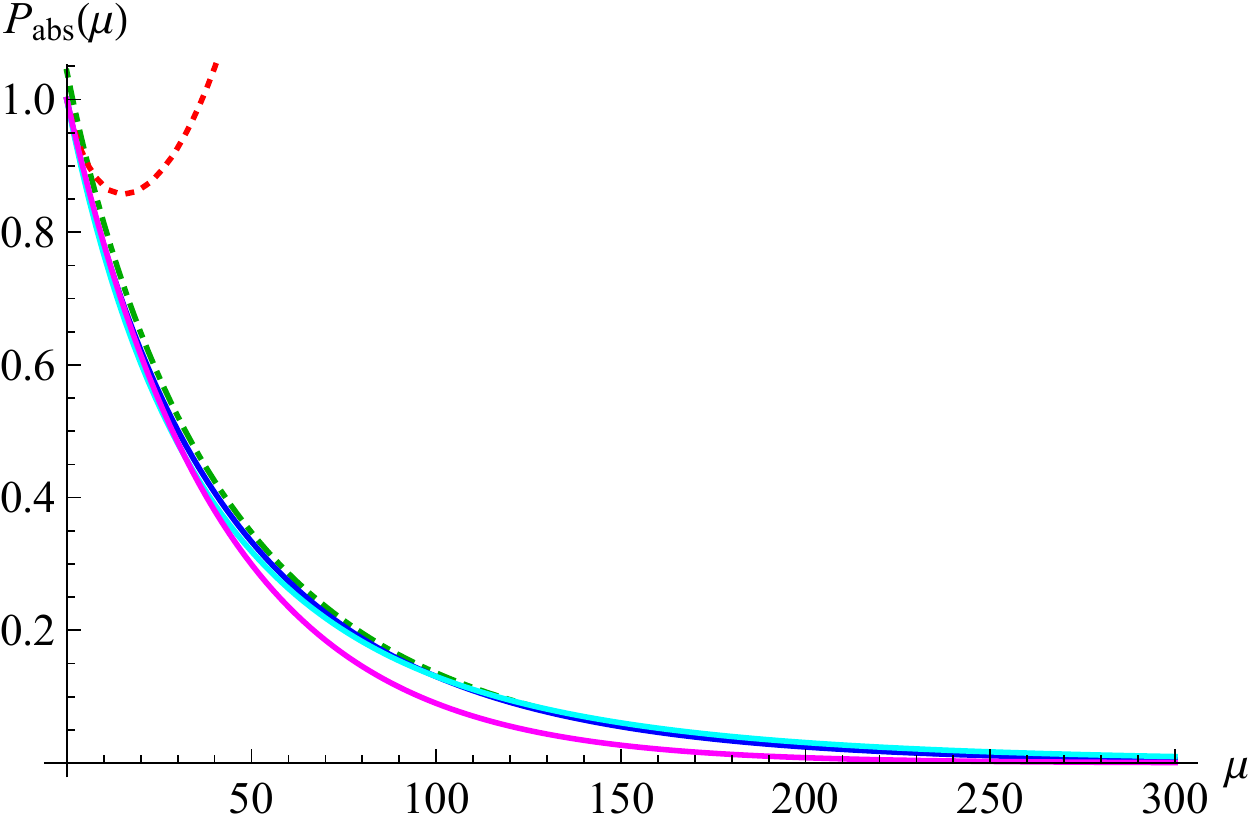}\hfill\fig{8.8cm}{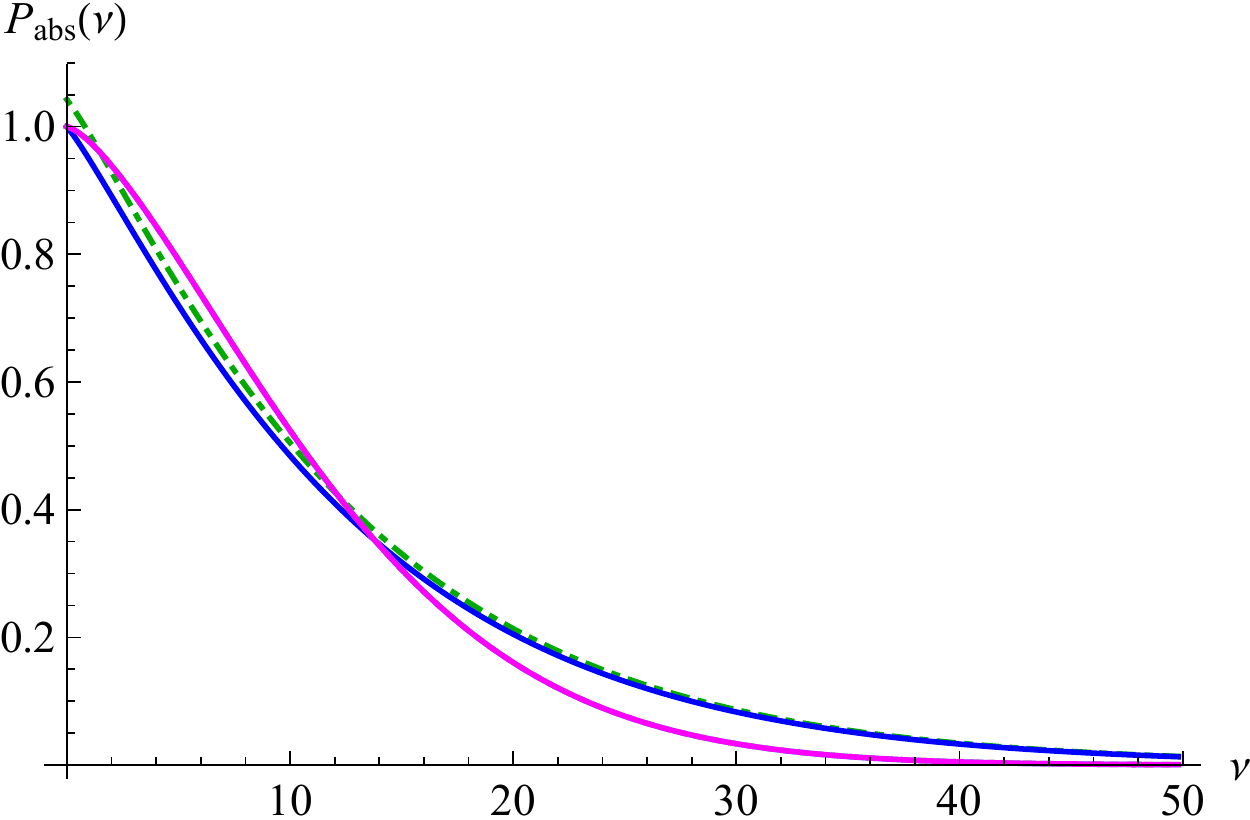}
\caption{Example for the absorption probability as a function of $\mu$ at $\nu=0$ (left), and $\nu$ at $\mu=0$ (right). In all cases $m=0.1$. The blue solid line represents the result obtained by a direct numerical integration of \Eq{P(y|m)-final}, and adjusting the overall normalisation at $\mu=\nu=0$ to 1; this has the advantage that the combination $\mu m^{\frac1H-1}$ appears naturally. The green dashed curve is the same, without adjustment of normalisation.  The red dotted curve (visible only on the left plot) is obtained using \Eq{P-abs-final1}. The magenta curve is obtained using  \Eq{P-abs-final2}. The cyan curve is from  \Eq{P-abs-final3}, and is identical to the magenta one  on the right plot.}
\end{figure*}
\end{widetext}
From the appearance of fractal powers of $m$ and $\nu$ in \Eq{97}, we suspect that both power series in $m \mu^{\frac{H}{1-H}}$ and    $m^{\frac1H-1} \mu$ might appear. 
While numerical simulations could decide which version is a better approximation, only higher-order calculations would be able to settle the question.

\subsection{Relation between the full propagator, first-passage times, and the distribution of the maximum}\label{s:remark}
In this section, we demonstrate how the probability densities of three different observables follow from the same scaling function. This shows how our result can be used to find the probability distribution of both running maxima and first-passage times for fBM with linear and non-linear drift. 

Let us start with the drift-free case, $\mu=\nu=0$.
\begin{enumerate} 
\item[(i)] In Ref.~\cite{WieseMajumdarRosso2010} was calculated 
$     \mathbb P_+(m,t)$, 
the {\em normalised} probability density to be at $m$, given $t$, when starting at $x_{0}$ close to 0 {(in \cite{WieseMajumdarRosso2010} this quantity is denoted $P_{+}(x,t)$ with $m=x$)}. While  $\mathbb P_+$ is a density in $m$, and thus should be denoted $  P_+$ (cf.~Tab.~\ref{t:notations}), it is the time derivative of a probability, see \Eq{55}. This can be seen  from its definition, 
\be
 \mathbb P_+(m,t)  := \frac{  P_+(m,t|x_0) }{\int_0^\infty \rmd m\, P_+(m,t|x_0) }\ ,
\ee
and the asymptotic expansion at small $x_0$, (see e.g.\ \cite{WieseMajumdarRosso2010}, appendix C)
\be
\int_0^\infty \rmd m\, P_+(m,t|x_0)  \sim x_0^{\frac1H-1}\ ,
\ee
which implies that $ \mathbb P_+(m,t) $ has dimension $1/$time. 

\item [(ii)]
Here we consider the probability density to be absorbed at time $t$ when starting at $m$.  
This is a first-passage time, with distribution 
$\mathbb P_{\rm first}(m,t)$. 

\item [(iii)]
Third, let the process start at 0, and consider the distribution of the max $m$, given a total time $t$, $  P_{\rm max}(m,t)$, denoted by $P^{T}_H(m)$ (with $t=T$) in 
Ref.~\cite{DelormeWiese2016}. \end{enumerate}
All three objects have a scaling form depending on the same variable 
$y= \frac{m}{\sqrt{2}t^H}$:
\bea\label{51}
\mathbb P_{\rm first}(m,t)
&=&  \frac {H}t \ca P_{\rm first}(y)\, 
\ ,
\\
\label{52}
\mathbb  P_{+}(m,t) 
&=&  \frac {H}t \ca P_{+}(y) 
\ ,
\\ 
\label{53}
 P_{\rm max}(m,t) 
 &=&  \frac1{\sqrt 2 T^{H}} \ca P_{\rm max}(y)
 \ .
\eea
The factors of $H$ and $\sqrt 2$ where chosen for later convenience. 
These objects are related.
Denote 
$\mathbf P_{{\rm surv}}(m,t)$ the probability   to start at $x=0$, and to survive in presence of an absorbing boundary at  $m$ up to time $t$. 
Note that $\mathbf P_{{\rm surv}}(m,t)$ is a probability, whereas $\mathbb P_{{\rm first}}(m,t)$, $\mathbb P_{{+}}(m,t)$, and $  P_{{\rm max}}(m,t)$ are densities, the first two in $t$, the latter in $m$. 
Then 
\bea\label{55}
\mathbb P_{+}(m,t)&=& \mathbb P_{\rm first}(m,t) = -\partial_{t } \mathbf P_{\rm surv}(m,t)\ ,~~~\\
\label{56a}
 P_{\rm max}(m,t) &=& \partial_{m } \mathbf P_{\rm surv}(m,t)\ .
\eea
Since $\mathbf P_{{\rm surv}}(m,t)$ is a probability, it is scale free, and  scaling implies that
\be\label{57}
\mathbf P_{{\rm surv}}(m,t) = \mathbf P_{{\rm surv}}\!\left(y=\frac{m}{\sqrt 2 t^{H}}\right)\ .
\ee
Putting together \Eqs{55}, \eq{56a} and \eq{57} proves \Eqs{51} to \eq{53}, with 
\bea
\ca P_{\rm first}(y) &=& \ca P_{+}(y) = y \mathbf P_{\rm surv}'(y) \\
\ca  P_{\rm max}(y) &=& \mathbf  P_{\rm surv}'(y)\ .
\eea
The scaling functions appearing are {\em almost} the same, differing by (innocent looking) factors of $t$ and $H$ and a (non-innocent looking) factor of $y$. 
However, when changing to the measure in $y$, all of them become {\it identical}.
The survival probability in absence of a drift is given in Eqs.~(63)-(64) of Ref.~\cite{DelormeWiese2016}.

Let us finally add drift. Then the survival probability  $\mathbf P_{\rm surv}(y,\tilde u,v)$ depends on three variables introduced in \Eqs{12}-\eq{15bis}, setting there $x\to m$. Since $\tilde u = m \mu^{\frac H{1-H}}$, and $v= \nu m$ are both constants multiplying $m$, we can write $\mathbf P_{\rm surv}(y,\tilde u,v)=\mathbf P_{\rm surv}(y,m)$.
Using \Eqs{55} and \eq{56a}, we  find 
\bea
 {\mathbb P}_{+}(m,t) &=&  {\mathbb P}_{\rm first}(m,t) = -\frac{\rmd}{\rmd t} \mathbf P_{\rm surv}(y,m) \nn\\
 &=& \frac{H}{t} \partial_{y}  \mathbf P_{\rm surv}(y,m) \ ,
 \\
 { P}_{\rm max}(m,t)  &=&  \frac{\rmd}{\rmd m} \mathbf P_{\rm surv}(y,m)\nn\\
 & =& \left[ \frac{y}m \partial_{y} + \partial_{m} \right] \mathbf P_{\rm surv}(y,m) \ .
\eea
Passing to the measure in $y$, we obtain
\bea
  {\ca P}_{+}(y,m)  &=&  {\ca P}_{\rm first}(y,m)  = y \partial_{y}  \mathbf P_{\rm surv}(y,m)\ ,\qquad \\
    {\cal P}_{\rm max}(y,m)  &=& \left[  \partial_{y} +\frac m y \partial_{m}\right]  \mathbf P_{\rm surv}(y,m)\ .
\eea
This set of equations allows us to express $\ca P_{\rm max}(y,m)$ as an integral over $\ca P_{+}(y,m) = \ca P_{\rm first}(y,m)$.

\subsection{Tail of the distribution}
\label{s:tails}
Piterbarg \cite{PiterbargBook2015} states (section 11.3, page 85) that for a fBm defined on the interval $[0,1]$, with $\left<x_{1}^{2}\right>=1$, in the limit of $u \to \infty$, 
\bea \label{19.0}
&&\!\!\! \mathbf {P}(\mbox{max}_{0\le t\le 1} \,x_t >u ) \nn\\
  &&  \simeq \mathbf\Psi(u) \times
  \left\{ 
\begin{array}{ccrcl}
   2 \ \ &,&  \quad H&=&1/2 \\
 1 \ &,& \quad   H&>&1/2\\   
  \displaystyle \frac{{\cal H}_{2H}}{2H} 2^{\frac 1 {2H}} u^{\frac1H-2}\ &,& ~~    H&<&1/2 \\
\end{array} \right.\ . \\
 &&\!\!\! \mathbf \Psi (u) := \frac{1}{ \sqrt{2\pi}u}   \exp\left(-\frac{u^2}2\right) \simeq \frac{1}{ \sqrt{2\pi}} \int_u^\infty \exp\left(-\frac{x^2}2\right) \rmd x\ .\nn\\
\eea 
The estimate for $H<1/2$ seems to contain misprints: We 
 find $\sigma(t):= \sqrt{\left< x_{t}^{2} \right>} = 1-H |1-t|$ (i.e.\ $H$ instead of $2H$). Rescaling $t-1\to (t-1) \times 2^{\frac1{2H}}$ gives $\sigma(t)  \to 1-H\times 2^{\frac1{2H}}\times |1-t|$, thus 
\be
 \mbox{\bf P}(\mbox{max}_{0\le t\le 1}\, x_t >u ) \simeq \frac{{\cal H}_{2H}}{2^{\frac 1 {2H} }H }  u^{\frac1H-2}\mathbf\Psi(u)\ , ~~    H<\frac12. 
\ee
Using the latter result,
taking a derivative w.r.t.\ $u$, and passing to the measure in $y$,  one obtains   ${\cal  P}(y) \equiv \ca P_{>}(y|m,\mu=\nu=0) \equiv {\cal  P}_{\rm max}(y)$ (in terms of our variable $y$), in the limit of large $y$, 
 \be \label{19.1}
{\cal  P}(y)\simeq \frac{\rme^{-\frac {y^2}2}}{ \sqrt{2\pi}}   \times
  \left\{ 
\begin{array}{ccrcl}
   2 \ \ &,&  \quad H&=&1/2 \\
 1 \ &,& \quad   H&>&1/2\\   
 \displaystyle
   \frac{{\cal H}_{2H}}{  2^{\frac 1 {2H}} H } y^{\frac1H-2}\ &,& ~~    H&<&1/2 \\
\end{array} \right.\ . \\
\ee
The Pickands constant ${{\cal H}_{2H}}$ has $\epsilon $-expansion  \cite{DelormeRossoWiese2017}
\be\label{Pickands}
{{\cal H}_{2H}} = 1-2 \gamma_{{\rm E}} \epsilon +{\cal O}(\epsilon)^{2}\ .
\ee
How is this consistent with \Eq{P(y|m)-final}?
Taylor-expanding the latter for large $y$ yields 
\bea\label{28}
{\cal  P} (y) \simeq  2  \frac{  \rme^{{-y^2/2}}} {\sqrt{ 2 \pi }}\Big\{ 1 &\!\!-\!\!& \Big[1+\gamma_{{\rm E}} +2\ln(y)+\ln(2) \Big] {\epsilon} \nn\\
&\!+\!& {\cal O}(\epsilon^{2}) +{o}(y^{0})\Big\}\ .
\eea
In Ref.~\cite{WieseMajumdarRosso2010} this was  interpreted as ${\cal  P} (y) \sim y^{-2 \epsilon} \rme^{-y^2/2}$. \Eq{19.1} shows that this interpretation is incorrect. For large $y$, our expansion is almost  the sum of the two contributions in \Eq{19.1} for $H\neq 1/2$,  
\bea\label{guess}
{\cal  P}(y) &\approx& \frac{\rme^{{-y^2/2}}} {\sqrt{2\pi }}\left[ 1+  \frac{{\cal H}_{2H}}{  2^{\frac 1 {2H}}H }  y^{\frac1H-2}  + ... \right]\nn \\
& \simeq&  2  \frac{  \rme^{{-y^2/2}}} {\sqrt{ 2 \pi }}\Big\{ 1  -  \Big[1+\gamma_{{\rm E}} +2\ln(y)-\ln(2) \Big] {\epsilon} \nn\\
&& ~~~~~~~~~~~~~~~~~~ + {\cal O}(\epsilon^{2})  +\ca O(y^{0})\Big\}\ .
\eea
Note the difference in sign for the $\ln(2)$ term  between Eqs.~\eq{28} and \eq{guess},   showing that the   guess \eq{guess} slightly underestimates the amplitude for $\epsilon<0$.

\section{Numerics}
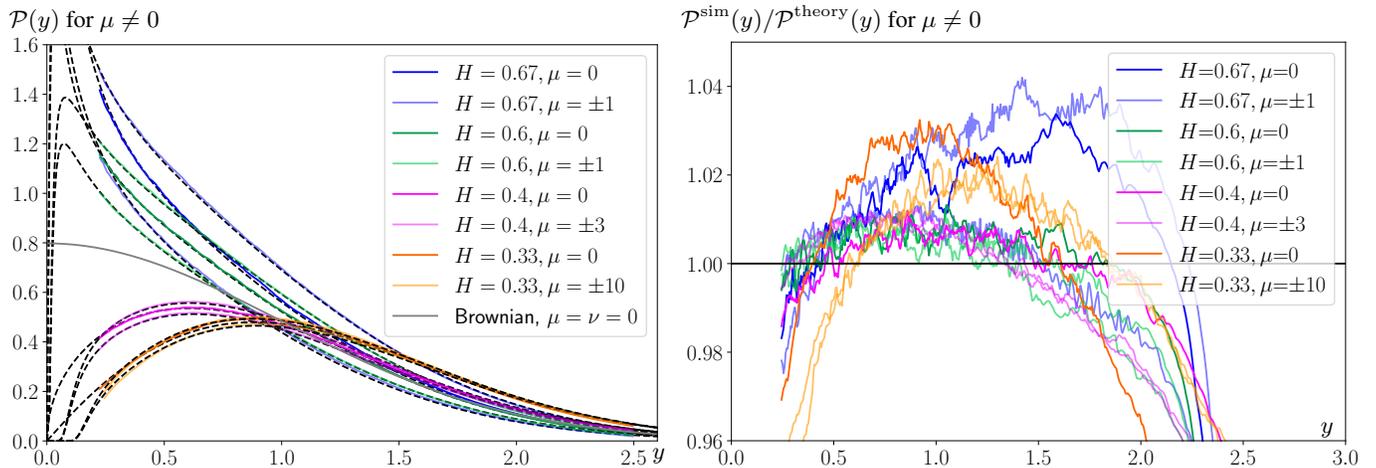
\begin{figure*}[t]
{\fboxsep0mm
\mbox{\setlength{\unitlength}{1cm}\begin{picture}(8.7,6.2)
\put(0,0){{\includegraphics[trim=50 30 63 45,clip,width=1\columnwidth]{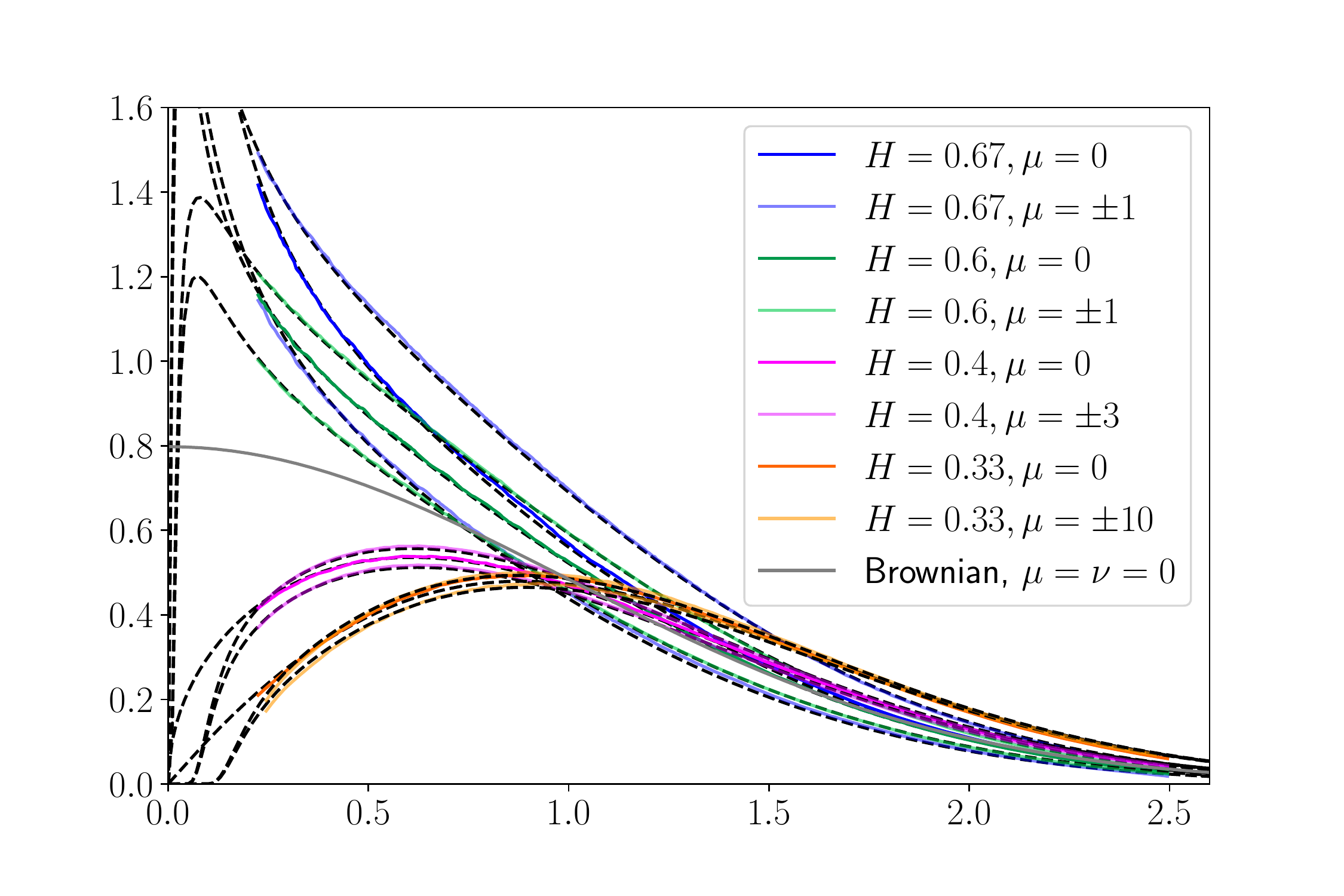}}}
\put(8.55,0.15){$y$}
\put(0,5.9){$\ca P(y)$ for $\mu\neq 0$}
\end{picture}}}
\hfill{\fboxsep0mm
\mbox{\setlength{\unitlength}{1cm}\begin{picture}(8.99,6.2)
\put(0,0){\mbox{\includegraphics[trim=40 30 55 49,clip,width=1.04\columnwidth]{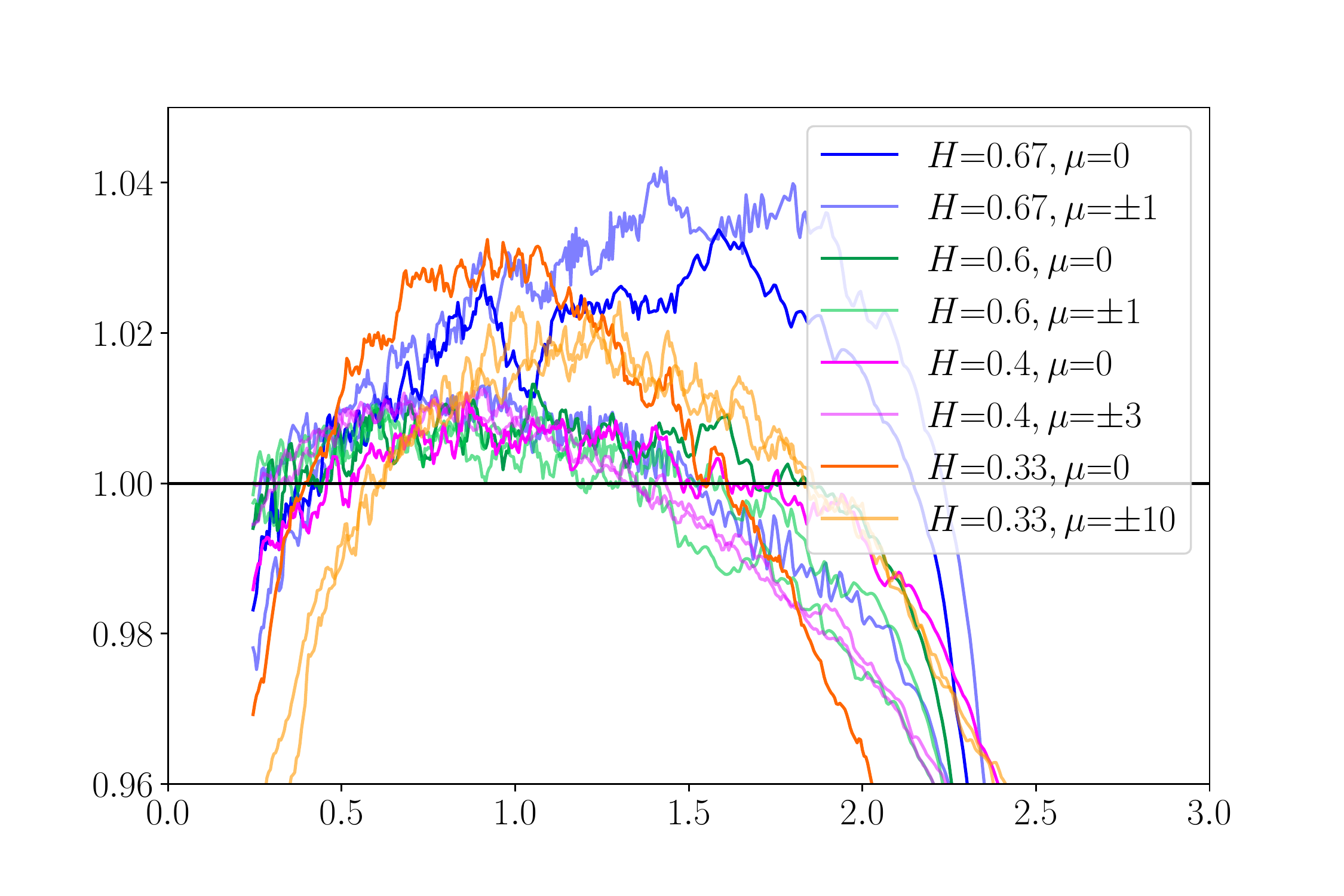}}}
\put(8.5,0.5){$y$}
\put(0,5.9){$\ca P^{\rm sim}(y)/{\cal P}^{\rm theory}(y)$ for $\mu \neq 0$}
\end{picture}}}
\caption{Left: First-passage time density $\ca P_{\rm first}(m,t)=\ca P(y)$ plotted as a function of $y$ as given in \Eq{y}. In order to increase the resolution of the plot, we use  overlapping bins with binsize  $5\times 10^5$, with $y$ increasing by  $10^5$ points for each bin;  $m=0.1$. For various values of $H$ and $\mu$, numerical simulations are compared to the theory. As can be seen on this plot, and on the ratio between simulations and theory to the right, the relative error is about $3\%$ at the extreme points. Note that neglecting   $\ca F_1(y)$ would lead   for $H=0.4/0.6$ to an error of $15\%$, and for $H=0.33/0.67$ to an error of $25\%$.}
\label{f:sim-lindrift}
\end{figure*}

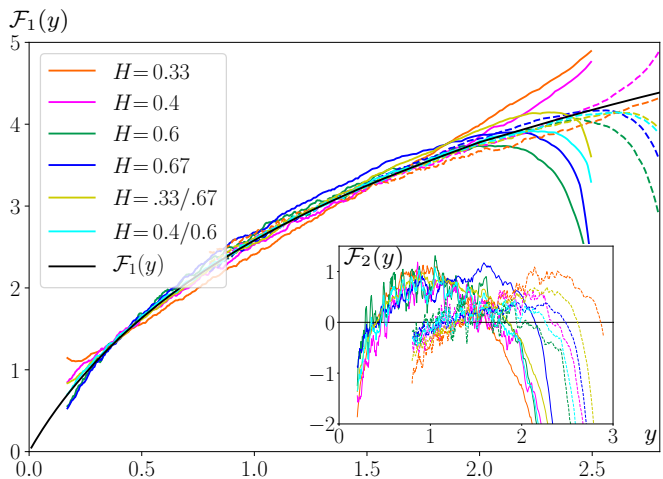
\begin{figure}[t]
\centerline{\fboxsep0mm
\mbox{\setlength{\unitlength}{1cm}\begin{picture}(8.65,6.3)
\put(0,0){{\includegraphics[trim=65 30 62 45,clip,width=1.006\columnwidth]{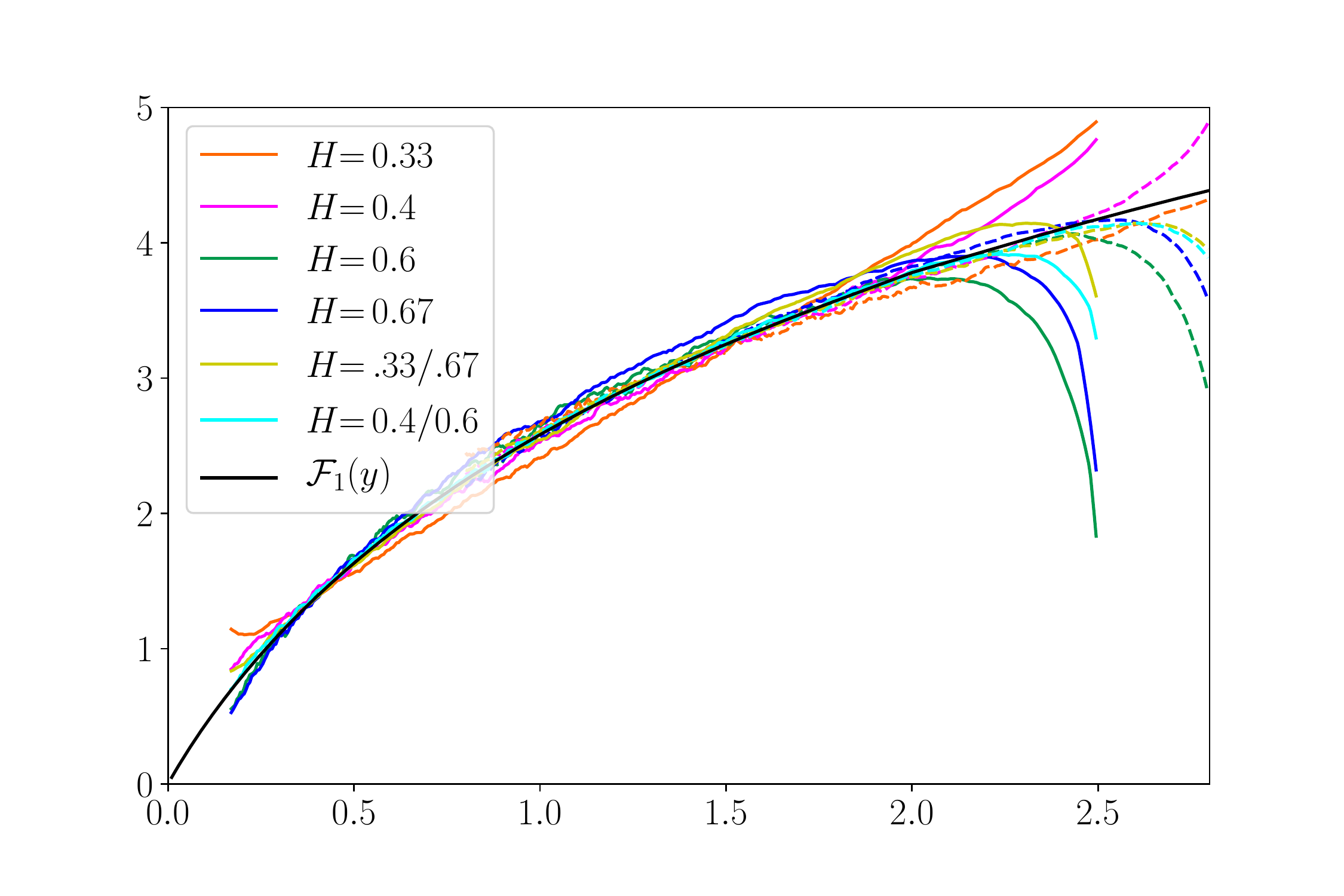}}}
\put(3.8,.45){\mbox{\includegraphics[trim=0 10 52 49,clip,width=0.5\columnwidth]{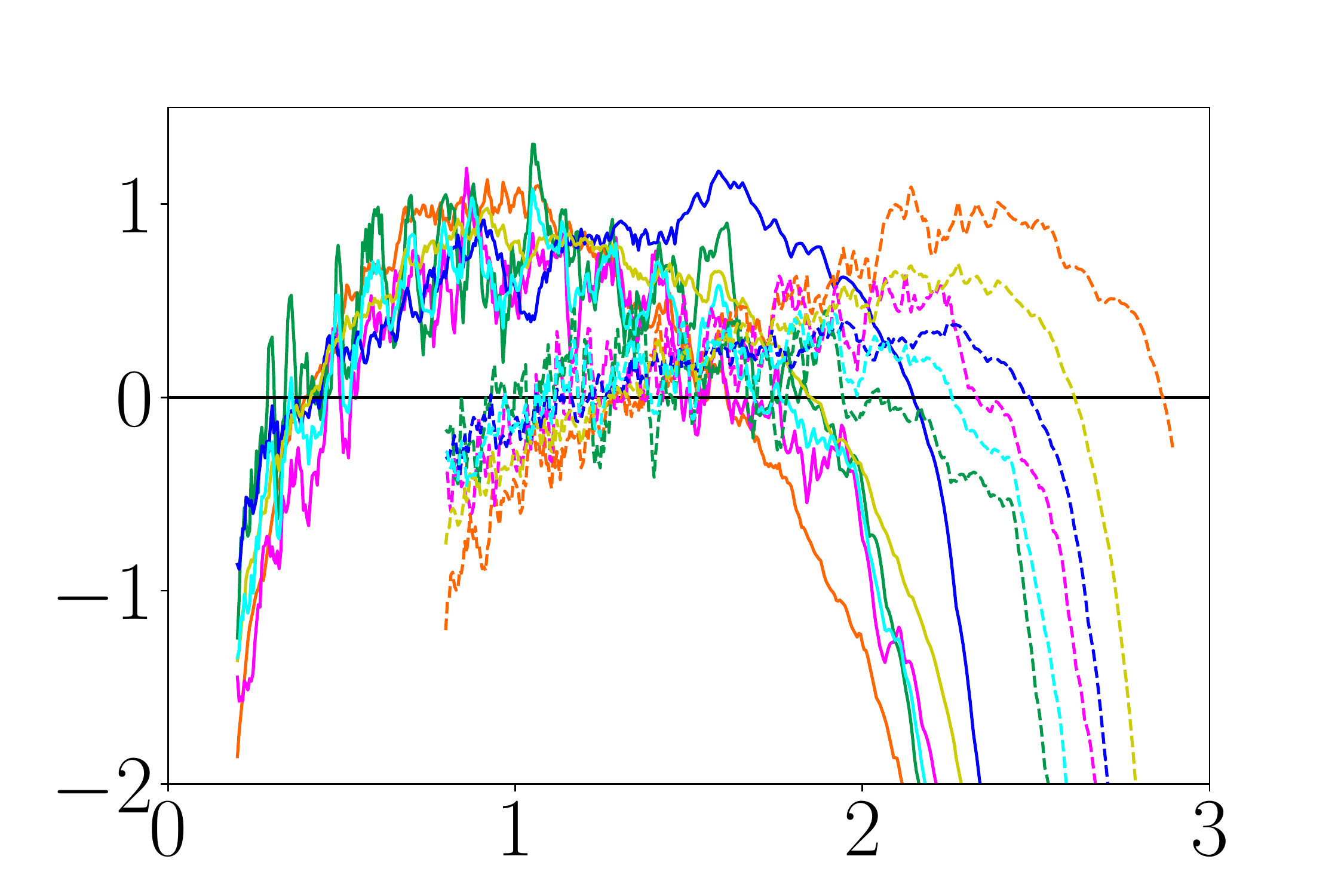}}}
\put(8.45,0.55){$y$}
\put(0,6.05){$\ca F_1(y)$}
\put(4.45,2.9){$ \ca F_2(y)$}
\end{picture}}}
\caption{Numerical estimate of $\ca F_{1}$. The black curve is the theoretical estimate \eq{F1}, followed by a number of estimates using \Eq{F1-est}. 
 Solid lines are for $m=0.1$, dashed ones for $m=1$. The symmetrised estimates \eq{F1-est-sym} are in olive/cyan.
The latter has minimal deviations from the theory. The  inset shows a numerical estimate for $\ca F_{2}(y)$, as given by Eqs.~\eq{F2-est} and \eq{F2-est-sym}. All curves are consistent, and let appear even the next-to-leading corrections. (Remind that changing the   normalization is equivalent to adding a constant to $\ca F_{1}(y)$ or $\ca F_{2}(y)$). The strong curve-down  for small and large $y$ are due to numerical problems.}
\label{f:F1-num}
\end{figure}
\label{s:Numerics}
\subsection{Simulation protocol}
\label{s:Simulation protocol}
Fractional Brownian motion can be simulated with the classical Davis-Harte (DH) algorithm \cite{DaviesHarte1987,DiekerPhD}, whose algorithmic complexity (execution time) scales with system size $N$ as $N \ln N$. Here we use the adaptive bisection algorithm introduced and explained in Refs.~\cite{WalterWiese2019a,WalterWiese2019b}. For $H=1/3$  its measured algorithmic complexity grows as $(\ln N)^3$, making it   about 5000 times faster, and 10000 times less memory consuming than DH for   an effective grid size of $N=2^{32}$.

To measure the functions $\ca F_{1}$, $\ca F_{\mu}$ and $\ca F_{\nu}$, which all depend on   $y$ only, we \begin{enumerate}
\item[(i)] generate a (drift free) fBm $x_t$ with $x_0=0$, of length $N$; the latter corresponds to a time $T=1$. 
\item[(ii)] add the drift terms   to yield $z_t = x_t+\mu t + \nu t^{2H}$
\item[(iii)] for given $m$, find the first time $t$, s.t.\ $z_t= m$
\item[(iv)] evaluate $y=\frac m{\sqrt{2}t^H}$; add a point to the histogram of $y$. 
\end{enumerate}
This histogram misses values of $t>T=1$, i.e.\ $y<\frac{m}{\sqrt {2}}$.

We checked the procedure for  Brownian motion (with $\nu\to0$), where
\be
\ca P(y|m,\mu) = \sqrt{\frac{2}{\pi }}\, \rme^{-\frac{ (\mu  m+2
   y^2 )^2}{8 y^2}} 
\ .
\ee
Note that this is a function of $y$ and $m\mu$ only, so that we can write
\be
\ca P(y|   m  \mu) = \sqrt\frac{2}\pi \rme^{-\frac{y^2}2}\times \rme^{-\frac {m \mu} 2}\rme^{-\frac{(m \mu)^2} {8 y^2}} \ .
\ee
For fBm, we measure  $\ca P(y|m,\mu,\nu)$, and then extract $\ca F_{1}$, $\ca F_{\mu}$ and $\ca F_{\nu}$. Firstly, 
\be\label{F1-est}
\ca F_{1}^{\epsilon }(y|m) := \frac1 \epsilon \ln \bigg( \ca P(y|m) y^{2-\frac1H}\rme^{\frac{y^{2}}2}\bigg) \bigg|_{\mu=\nu=0}
\ee
and $\ca F_{1}^{\epsilon }(y|m)  =\ca F_{1}(y) +\ca O(\epsilon^2)  $. The following combination is more precise, since terms even in $\epsilon$ cancel, 
\be\label{F1-est-sym}
\ca F_{1}^{\epsilon,\rm sym}(y|m) = \frac12 \Big[ \ca F_{1}^{\epsilon }(y|m)+ \ca F_{1}^{-\epsilon }(y|m) \Big]+\ca O(\epsilon^{ 2})
\ .
\ee
The second-order correction can be estimated as
\be\label{F2-est}
\ca F_{2}^{\epsilon}(y|m) := \frac1{\epsilon} \left[\ca F_{1}^{\epsilon }(y|m)- \ca F_{1} (y|m) \right]+\ca O(\epsilon)
\ .
\ee
Its symmetrised version again suppresses subleading corrections, 
\be\label{F2-est-sym}
\ca F_{2}^{\epsilon,\rm sym}(y|m) := \frac1{2\epsilon} \Big[\ca F_{1}^{\epsilon }(y|m)-\ca F_{1}^{-\epsilon }(y|m) \Big]+\ca O(\epsilon^2)
\ .
\ee
The third order correction can be extracted as 
\bea\label{F3}
\ca F_{3}^{\epsilon}(y|m) &:=& \frac1{2\epsilon^{2}} \left[\ca F_{1}^{\epsilon }(y|m)+\ca F_{1}^{-\epsilon }(y|m)- 2\ca F_{1} (y|m) \right]\nn\\
&& +\ca O(\epsilon)\ .
\eea
For the remaining functions $\ca F_{\mu}$ and $\ca F_{\nu}$, we can employ similar formulas; we have to decide how to subtract $\ca F_{1}$, numerically from the simulation, or analytically, i.e. by supplying numerically or analytically the denominator in 
\bea
\ca F_{\mu}^{\epsilon }(y|m,\mu) &:=&- \frac1 \epsilon \bigg[  \ln \bigg( \frac{\ca P(y|m,\mu,\nu =0) }{\ca P(y|m,\mu=\nu=0)} \bigg)  \times 
\frac{y^{{ - 2 \epsilon}}}{\mu m^{{\frac 1H-1}}} \nn
\\
&& 
\qquad +\frac12 +\frac {\mu}{4}\left(\frac{m}{2}\right)^{\!\!\frac1H-1}y^{{3-\frac 5{2H}}}\bigg]
\ ,
\\
\ca F_{\nu}^{\epsilon }(y|m) &:=&- \frac1 \epsilon \bigg[  \ln \bigg( \frac{\ca P(y|m,\mu=0,\nu ) }{\ca P(y|m,\mu=\nu=0)} \bigg)  \times 
\frac{y^{- 2\epsilon}}{\nu m}\nn
\\
&& 
 \qquad+\frac12 +\frac {\nu m}{8} y^{-\epsilon - 2}\bigg]\ .
\label{Fnu-sim-v1}
\eea
We can also work symmetrically
\be\label{Fmu-estimate-sym}
\ca F_{\mu}^{\epsilon }(y|m) :=- \frac1 \epsilon \!\left[ \ln\! \bigg( \frac{\ca P(y|m,\mu,\nu {=}0) }{\ca P(y|m,-\mu,\nu{=}0)} \bigg)   
\frac{y^{{ - 2 \epsilon}}}{2\mu m^{{\frac 1H-1}}} {+}\frac12  \right]
 .
\ee 
\be
\label{Fnu-estimate-sym}
\ca F_{\nu}^{\epsilon }(y|m) :=- \frac1 \epsilon \left[  \ln \bigg( \frac{\ca P(y|m,\mu{=}0,\nu ) }{\ca P(y|m,\mu{=}0,-\nu)} \bigg)   
\frac{y^{- 2\epsilon}}{2\nu m} +\frac12 \right]
\ .
\ee

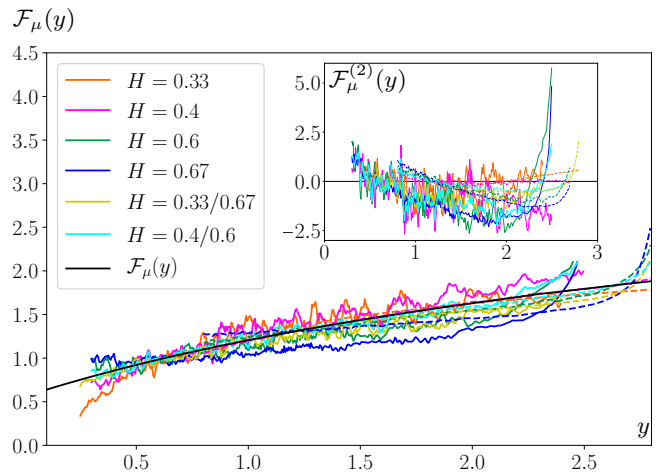
\begin{figure}[t]
\centerline{\fboxsep0mm
\mbox{\setlength{\unitlength}{1cm}\begin{picture}(8.65,6.2)
\put(0,0){\mbox{\includegraphics[trim=53 30 55 45,clip,width=1\columnwidth]{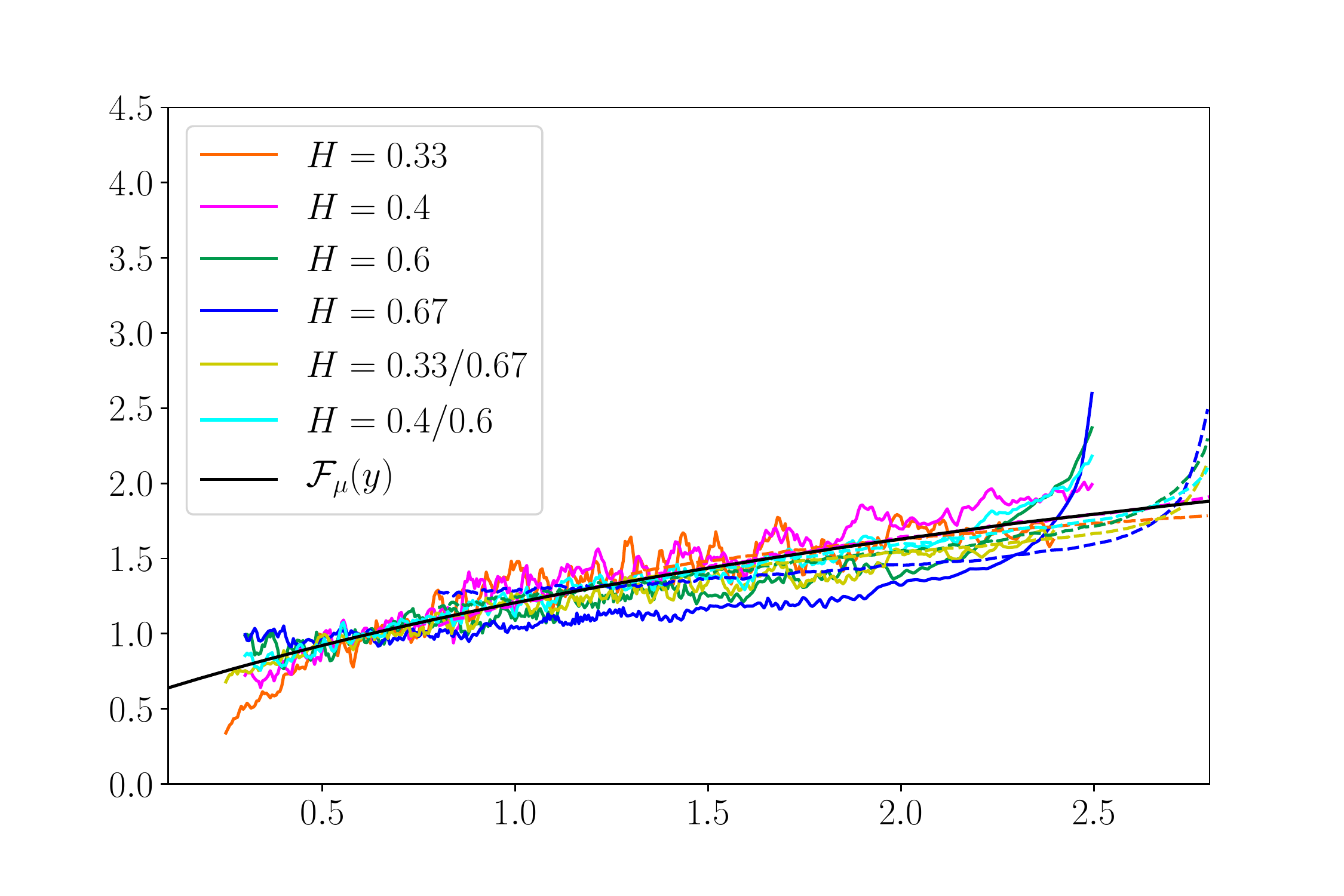}}}
\put(3.6,2.85){\mbox{\includegraphics[trim=6 18 44 40,clip,width=0.5\columnwidth]{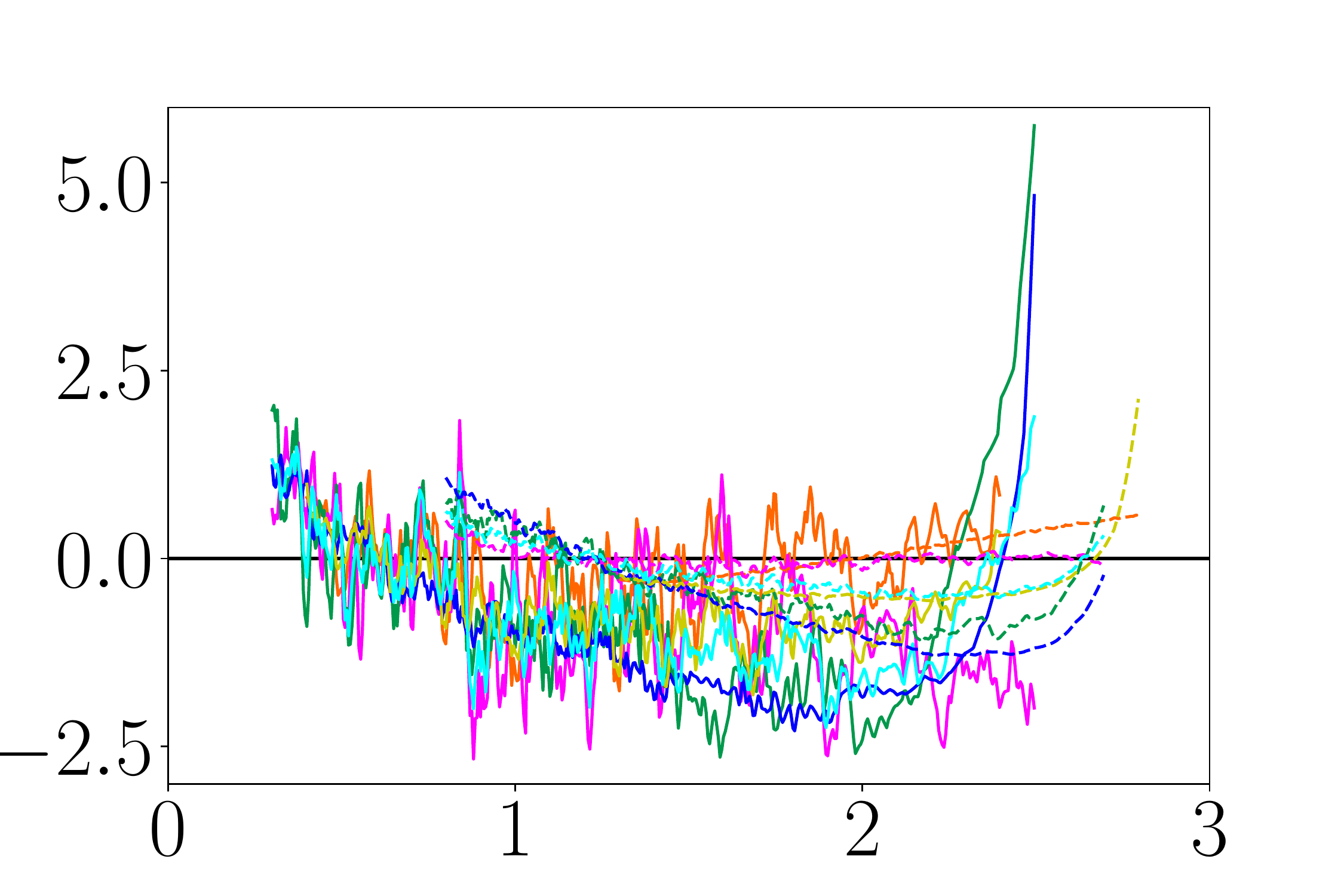}}}
\put(8.3,0.55){$y$}
\put(0,5.96){$\ca F_{\mu}(y)$}
\put(4.2,5.15){$\ca F_{\mu}^{{(2)}}(y)$}
\end{picture}}}
\caption{Numerical estimate of $\ca F_{\mu}$. The black curve is the theoretical result \eq{Fmu}. The colored curves are obtained using \Eq{Fmu-estimate-sym} with $\mu = \pm 1$ for $H=0.6$ and $H=0.67$, and $\mu = \pm 3$ for $H=0.33$ and $H=0.4$.   Solid lines are for $m=0.1$, dashed ones for $m=1$. The symmetrised estimates \eq{F1-est-sym} are in olive/cyan. The cyan curve using the equivalent of \Eq{F1-est-sym} with $H=0.4/0.6$ is our best numerical estimate of $\ca F_\mu(y).$ The inset shows the estimated second-order correction, analogous to Eqs.~\eq{F2-est}-\eq{F2-est-sym}.}
\label{f:Fmu-num}
\end{figure}Finally, a more precise estimate of the theoretical curves is given by symmetrizing results for the same $|\epsilon|$, using the analogue of \Eq{F1-est-sym}.

%

\begin{figure*}[thb]
{\fboxsep0mm
\mbox{\setlength{\unitlength}{1cm}\begin{picture}(8.99,6.1)
\put(0,0){\mbox{\includegraphics[trim=50 30 62 45,clip,width=1.\columnwidth]{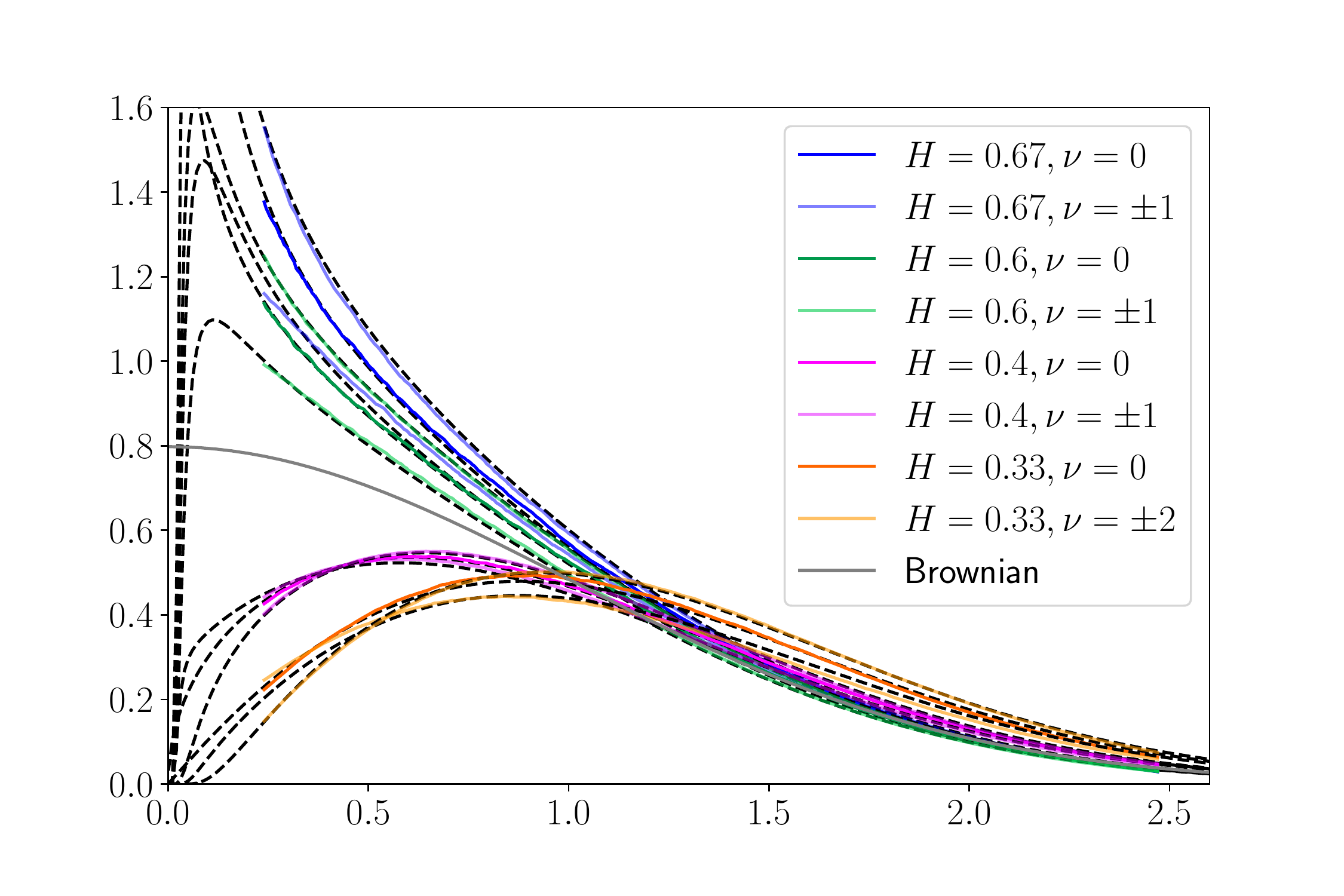}}}
\put(8.5,0.2){$y$}
\put(0,5.9){$\ca P(y)$}
\end{picture}}}
\hfill{\fboxsep0mm
\mbox{\setlength{\unitlength}{1cm}\begin{picture}(8.8,6.2)
\put(0,0){\mbox{\includegraphics[trim=45 33 63 46,clip,width=1.02\columnwidth]{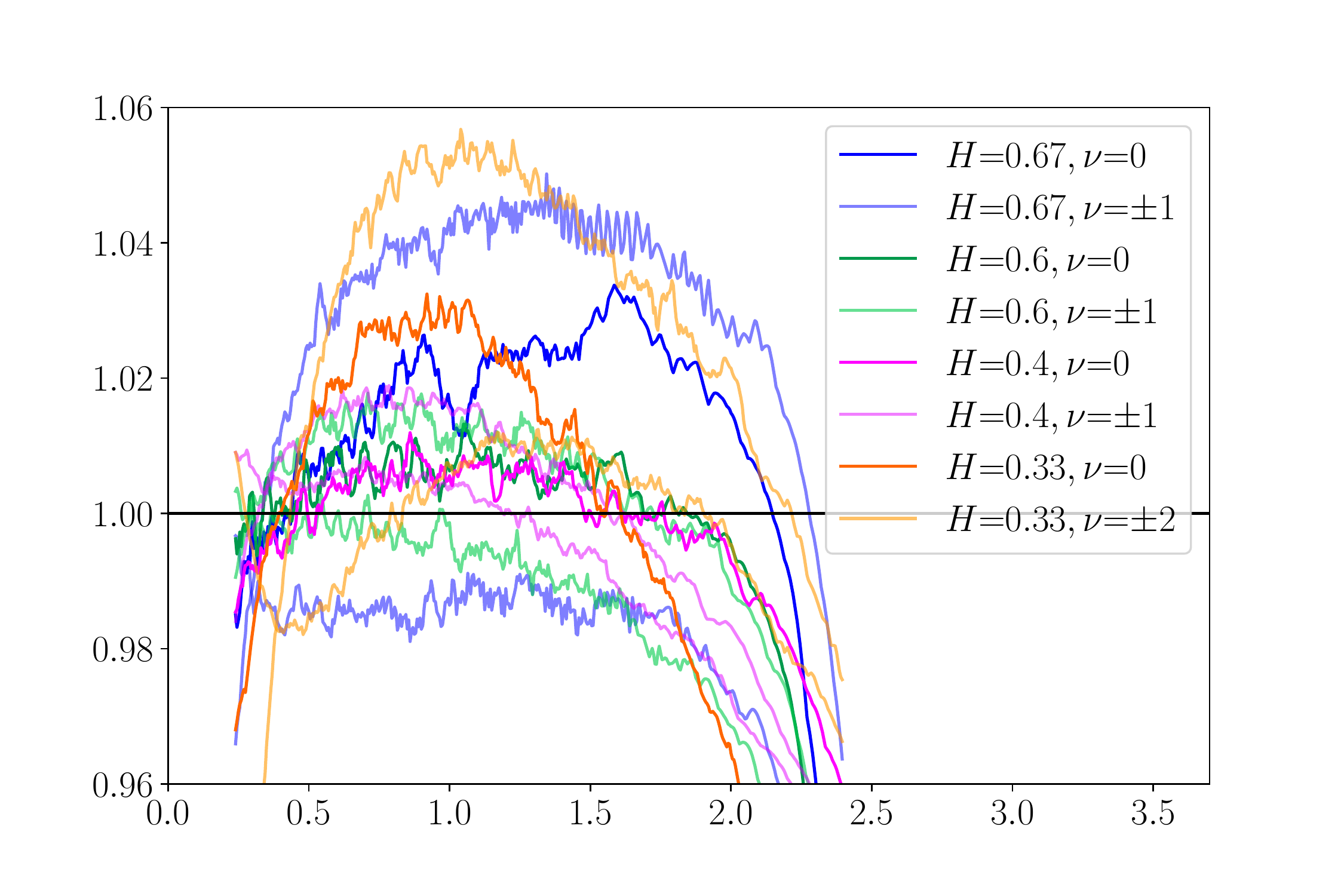}}}
\put(8.4,0.55){$y$}
\put(0,5.9){$\ca P^{\rm sim}(y)/{\cal P}^{\rm theory}(y)$ for $\nu \neq0$}
\end{picture}}}
\caption{Left: first-passage-time density plotted with overlapping bins as in Fig.~\ref{f:sim-lindrift} for various values of $H$ and non-linear  drift $\nu$ compared to the theory   given in \Eq{P(y|m)-final}. Right: Ratio of simulation and theoretical values.}
\label{f:sim-nonlindrift}
\end{figure*}
\begin{figure*}[thb]
\centerline{\fboxsep0mm
\mbox{\setlength{\unitlength}{1cm}\begin{picture}(8.65,6.5)
\put(0,0){\mbox{\includegraphics[trim=50 30 63 50,clip,width=1\columnwidth]{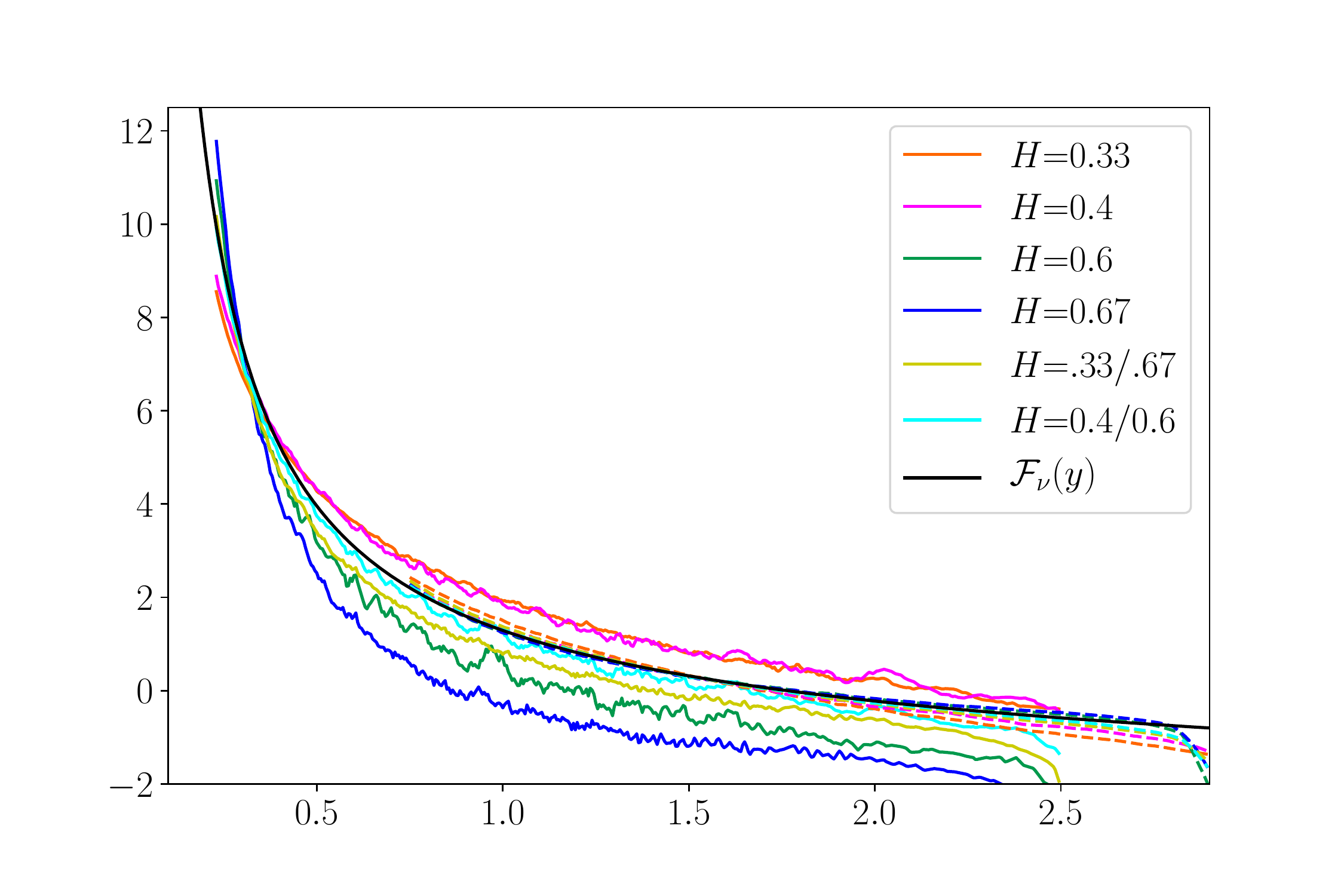}}}
\put(8.4,0.1){$y$}
\put(0,5.85){$\ca F_{\nu}(y)$}
\put(1.7,2.76){\mbox{\includegraphics[trim=23 18 64 40,clip,width=0.5\columnwidth]{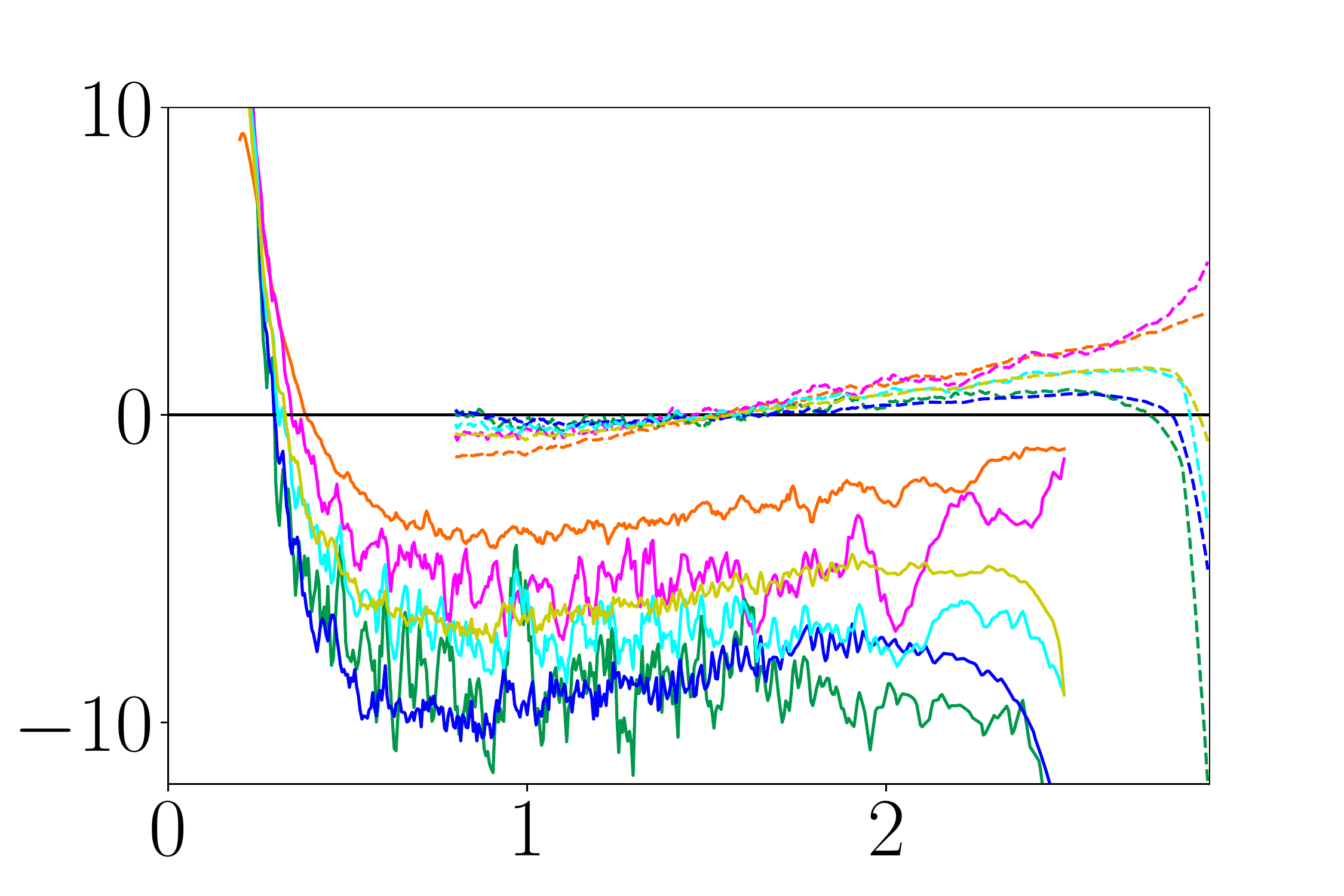}}}
\put(2.3,5.2){$\ca F_{\nu}^{{(2)}}(y)$}
\end{picture}}\hfill
\mbox{\setlength{\unitlength}{1cm}\begin{picture}(8.65,6.5)
\put(0,0){\mbox{\includegraphics[trim=50 30 63 50,clip,width=1\columnwidth]{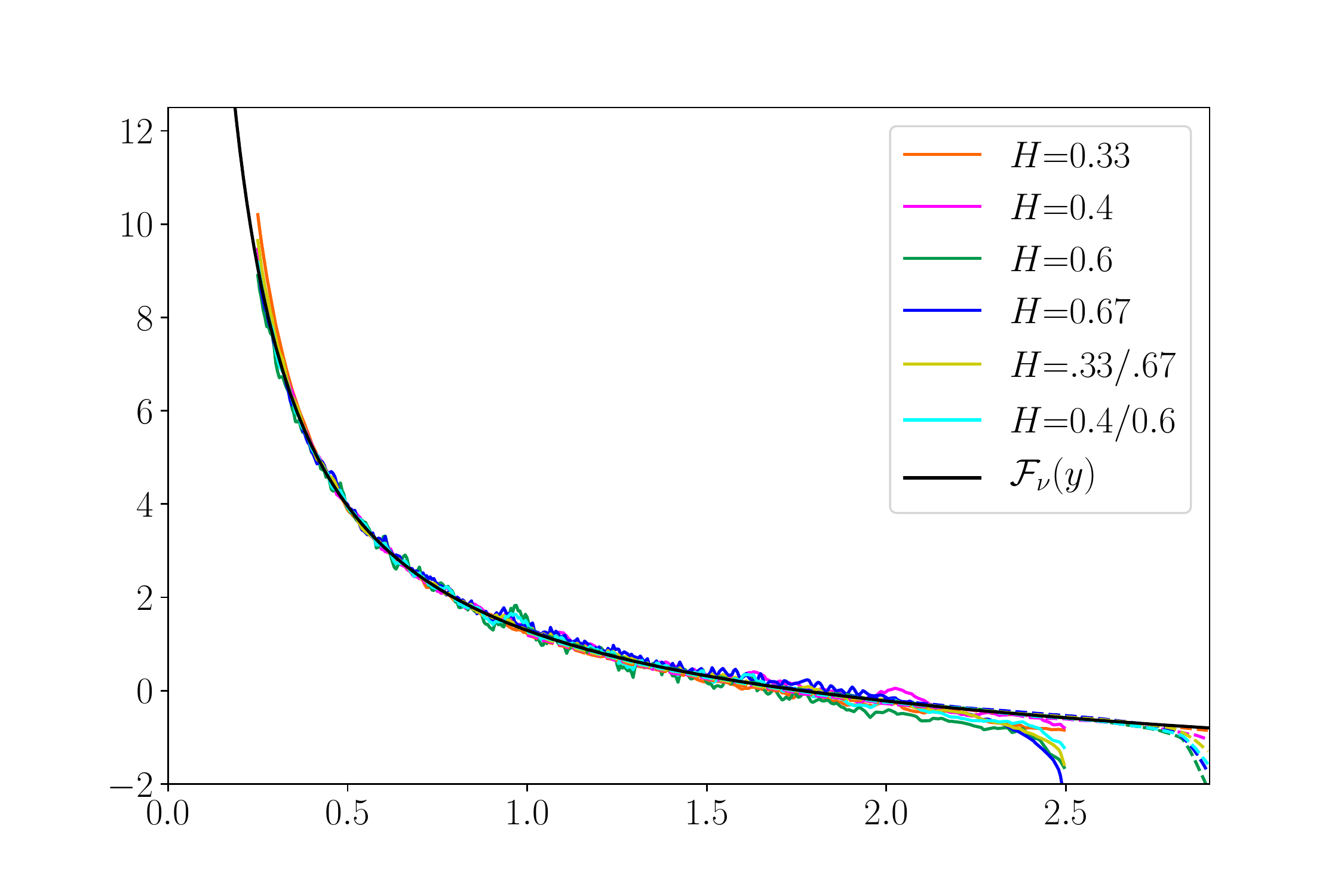}}}
\put(8.4,0.15){$y$}
\put(0,5.85){$\ca F_{\nu}(y)$}
\end{picture}}}
\caption{Left: Numerical estimate of $\ca F_{\nu}$, using \Eq{Fnu-estimate-sym}. The black curve is the theoretical prediction \eq{Fnu}.  The colored curves are simulation results using \Eq{Fnu-estimate-sym}.   Solid lines are for $m=0.1$, dashed ones for $m=1$. The cyan and olive curves are the symmetrised results using the equivalent of \Eq{F1-est-sym} for $H=0.4/0.6$ (cyan) and $H=0.33/0.67$ (olive).
The former one is the best numerical estimate of the theory, and very close to the latter. 
The  inset shows the estimated second-order corrections, analogous to Eqs.~\eq{F2-est}-\eq{F2-est-sym}.
There seem to be non-negligible  corrections of order three. 
An almost perfect data collapse can be obtained for $m=0.1$ as $
\epsilon \ca F^{{\epsilon}}_{{\nu}}(y)\simeq  \ca F_{\nu}(y) \epsilon +(2 y^{-2} -4 y^{-1} -6+y) \epsilon^{2}+(3y-20) \epsilon^{3}$, and for $m=1$ as 
 $
\epsilon \ca F^{{\epsilon}}_{{\nu}}(y)\simeq  \ca F_{\nu}(y) \epsilon +  (y-1.7) (1.5\epsilon^{2}-6\epsilon^{3})$,
 see right figure. Since  extrapolation problems mentioned around \Eq{Pade} become important for small $y$, this estimate is  intended as a  fit only, to show that the scatter on the left plot is consistent with higher-order corrections.   
}
\label{f:Fnu-num}
\end{figure*}

Below, we measure the three scaling functions $\ca F_{1}$, $\ca F_{\mu}$ and $\ca F_{\nu}$ for $H=0.33$, using our recently introduced adaptive-bisection   algorithm  \cite{WalterWiese2019a,WalterWiese2019b}. 
The latter starts out with an initial coarse grid of size $2^{g}$, which is then recursively refined up to a final   gridsize of $2^{g+G}$. It gains its efficiency by only sampling {\em necessary points}, i.e.\ those close to the target.  

The optimal  values of $g$ and $G$ depend on $H$. We run simulations with the following choices:   $H=0.33$ ($g=8$, $G=18$), $H=0.4$ ($g=10$, $G=14$), $H=0.6$ ($g=8$, $G=8$), and $H=0.67$ ($g=8$, $G=6$).  Thanks to the adaptive bisection algorithm,  we can maintain a resolution in $x$ of $10^{-3}$, with about 25 million samples at $H=0.33$, $H=0.6$ and $H=0.67$, and twice as much for $H=0.4$. As we will see below, this allows us to precisely validate our analytical predictions. 


\subsection{Simulation results}

\label{s:Simulation results}
We show simulation results on Figs.~\ref{f:sim-lindrift} to  \ref{f:Fnu-num}. First, on figure \ref{f:sim-lindrift} (left), we present results for the first-passage probability $\ca P(y|m,\mu,\nu=0)$,  using $m=0.1$. The numerical results (in color) are  compared to the predictions from \Eq{P(y|m)-final}. One sees that theory and simulations are in good quantitative agreement. This comparison is made more precise by plotting the ratio between simulation and theory on the right of    Fig.~\ref{f:sim-lindrift}. 

The function $\ca F_1(y)$ is extracted on Fig.~\ref{f:F1-num}.  We show simulations for $m=0.1$ (colored solid lines), and   $m=1$ (colored dashed lines). The theoretical result \eq{F1} agrees    with numerical simulations for all $H$,  at both values of $m$. Using the symmetrized form \eq{F1-est-sym} with $H=0.4/0.6$ shows a particularly good agreement. It  allows us to extract the subleading correction via \Eqs{F2-est} and \eq{F2-est-sym}. 
This is shown in the inset of Fig.~\ref{f:F1-num}; again the   symmetrized estimate is the most precise. 
 Note  that the second-order correction is rather sensitive to the choice of $m$; more effort would be needed to estimate it properly.
Also note that adding a constant to  $\ca F_{1}(y)$ is equivalent to an overall change in normalization, thus one should concentrate   on the shape of the cuves.

Using the data presented on Fig.~\ref{f:sim-lindrift},  Fig.~\ref{f:Fmu-num} shows the order-$\epsilon$ correction $\ca F_\mu$ extracted via \Eq{Fmu-estimate-sym}. The symmetrized estimate is rather close to the analytical result. 
The inset estimates the subleading correction.  Again, estimates for $m=0.1$ (dashed lines) and $m=1$ (solid lines) are consistent, and a proper measure of the second-order correction would demand a higher numerical precision. 

The results for non-linear drift $\nu$ are presented on Fig.~\ref{f:sim-nonlindrift}, starting with the probability distribution $\ca P(y|m) $ (left), followed by the ratio between simulation and theory on the right,  using $m=0.1$. The agreement is again  good. From these data is extracted the function $\ca F_\nu(y)$ defined in \Eq{Fnu}, see Fig.~\ref{f:Fnu-num}. 
Note that    $\ca F_\nu(y)$    is much larger than $\ca F_\mu(y)$  (Fig.~\ref{f:Fmu-num}),  and diverges for small $y$. 
The     subleading corrections to $\ca F_\nu(y)$ are not negligible,   {\em seemingly} $m$-dependent, and estimated as well, allowing us to collapse all measured estimates on the theoretical curve.  

In summary, we have measured all   scaling functions with good to excellent precision, ensuring that the analytical results are correct.

\section{Conclusion}
\label{s:Conclusion}
In this article, we gave analytical results for fractional Brownican motion, both with a linear and a non-linear drift. Thanks to a novel simulation algorithm, we were able to verify the analytical predictions with grid sizes up to $N=2 ^{28}$, leading to a precise validation of our results. 

Our predictions to first order in $H-1/2$ are precise, and many samples of very large systems are needed to see statistically significant deviations. We therefore hope that our formulas will find application in the analysis of data, as e.g.\ the stock market.

Another interesting question is how a trajectory   depends on its history, i.e.\ prior knowledge of the process. We   obtained analytical results also in this case, and will come back with its numerical validation in future work.

Our study can be generalised in other directions, as e.g.\ making the variance a stochastic process, as in \cite{ComteRenault1998} or in the  rough-volatility model of Ref.~\cite{GatheralJaissonRosenbaum2018}, which both use fBm in their modelling.

\medskip

\centerline{\bf Acknowledgements}

\medskip

It is a pleasure to thank J.P.~Bouchaud and  F.~Gorokhovik for discussions, G.~Pruessner for help with the implementation, and M.T.\ Jaekel and A.\ Thomas for support with the cluster. B.W.\ thanks LPTENS and LPENS for hospitality.

\appendix

\newcommand{\doi}[2]{\href{http://dx.doi.org/#1}{#2}}
\newcommand{\link}[2]{\href{http://#1}{#2}}
\newcommand{\arxiv}[1]{\href{http://arxiv.org/abs/#1}{#1}}


%

\ifx\doi\undefined
\providecommand{\doi}[2]{\href{http://dx.doi.org/#1}{#2}}
\else
\renewcommand{\doi}[2]{\href{http://dx.doi.org/#1}{#2}}
\fi
\providecommand{\link}[2]{\href{#1}{#2}}
\providecommand{\arxiv}[1]{\href{http://arxiv.org/abs/#1}{#1}}
\providecommand{\hal}[1]{\href{https://hal.archives-ouvertes.fr/hal-#1}{hal-#1}}
\providecommand{\mrnumber}[1]{\href{https://mathscinet.ams.org/mathscinet/search/publdoc.html?pg1=MR&s1=#1&loc=fromreflist}{MR#1}}

\appendix

\small
\tableofcontents
\end{document}